%% file: paper.tex
\documentclass[natbib,nocopyrightspace]{sigplanconf}

\usepackage{alltt}
\usepackage{mathptmx}
\usepackage{courier}
\usepackage[scaled=.92]{helvet} 
\usepackage[inline]{enumitem}  
\setlist[enumerate]{label=(\arabic*)} 
\usepackage{graphicx}
\graphicspath{{./figs/}}
\usepackage{setspace}
\usepackage{subfigure}
\usepackage{xspace}
\usepackage{booktabs}
\usepackage[hyphens]{url}  

\usepackage{todonotes}

\DeclareRobustCommand{\checked}[1]{#1\xspace}
\DeclareRobustCommand{\opinion}[1]{#1\xspace}

\begin{document}

\conferenceinfo{~} {~}
\CopyrightYear{2016}
\copyrightdata{2016}

\title{Author Growth Outstrips Publication Growth in Computer Science and
  Publication Quality Correlates with Collaboration}

\authorinfo{Stephen M. Blackburn}
           {Australian National University}
           {steve.blackburn@anu.edu.au}

\authorinfo{Kathryn S. McKinley}
           {Google}
           {ksmckinley@google.com}

\authorinfo{Lexing Xie}
           {Australian National University}
           {lexing.xie@anu.edu.au}

\maketitle


\section{Summary}

Although the computer science community successfully harnessed
exponential increases in computer performance to drive societal and
economic change, the exponential growth in publications is proving
harder to accommodate.  To gain a deeper understanding of publication
growth and inform how the computer science community should handle
this growth, we analyzed publication practices from several
perspectives: ACM sponsored publications in the ACM Digital
Library as a whole; subdisciplines captured by ACM’s Special Interest Groups
(SIGs); ten top conferences; institutions; four top U.S. departments;
authors; faculty; and PhDs between 1990 and 2012.  ACM publishes a
large fraction of all computer science research.

We first summarize how we believe our main findings inform
(1) expectations on publication growth, (2) how to distinguish research
quality from output quantity; and (3) the evaluation of individual
researchers.  We then further motivate the study of computer science
publication practices and describe our methodology and results in detail.

\begin{figure}
  \includegraphics[width=.9\columnwidth]{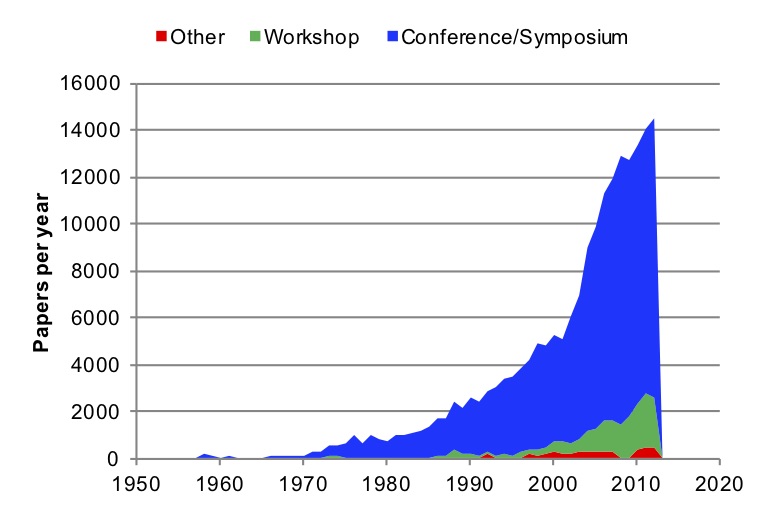}

\caption{Publications per year in the ACM DL experienced
  exponential growth of 9.3\% per annum between 1990 and 2012.}
\label{fig:acm-pubs}
\includegraphics[width=.9\columnwidth]{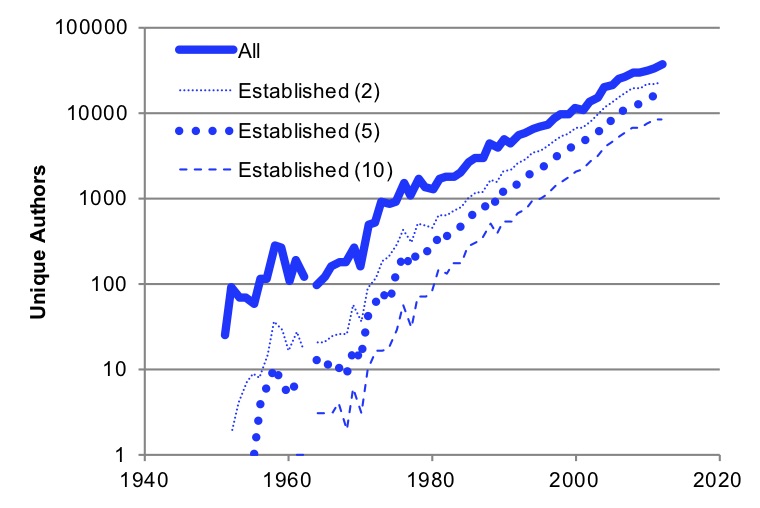}
\caption{Unique authors per year grew at 10.6\%
  per annum; faster than the growth in publications. Established authors with publications spanning $(N_e)$
  years grew at this same rate.}
\label{fig:acm-unique}
\includegraphics[width=.9\columnwidth]{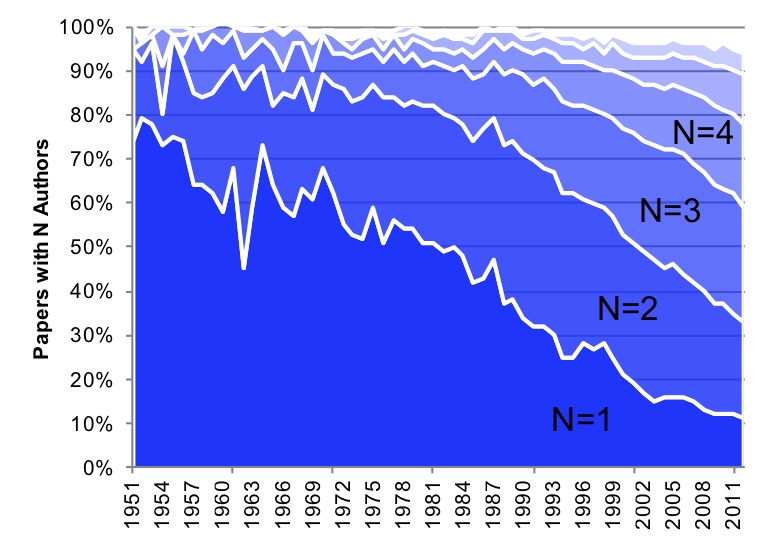}
\caption{Authors collaborate far more than before.   Single-author papers have become the exception rather than the rule.}
\label{fig:acm-papers-with-n-authors}
\label{fig:acm-percentile-avg-authors}
\includegraphics[width=.9\columnwidth]{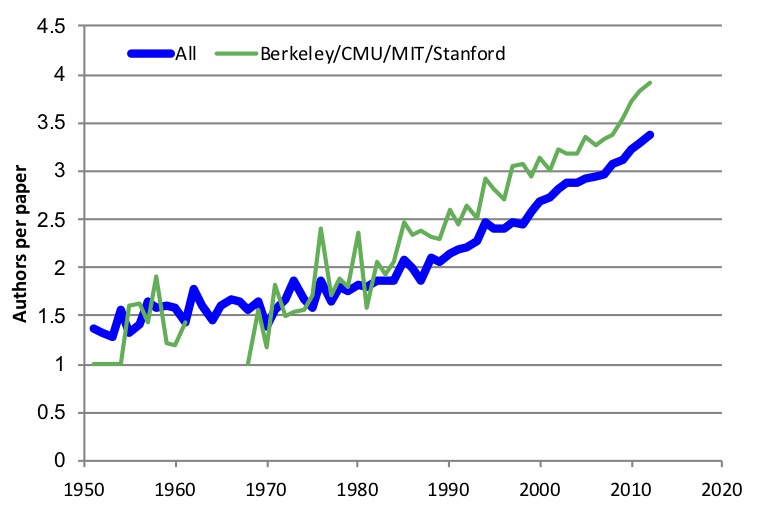}
\caption{Average authors per paper increased to 3.4 for the field and to
  3.9 at four top U.S. departments.}
\label{fig:acm-avg-authors}
\end{figure}

\textbf{(1) The number of computer science publications and researchers are
growing exponentially.} Figure~\ref{fig:acm-pubs} shows that computer
science publications experienced exponential growth (9.3\% per annum),
doubling every eight years. Figure~\ref{fig:acm-unique} shows that the
number of computer science authors increased even more than
publications. We distinguish \emph{established} researchers, whose
publications span five or more years ($N_e=5$), with authors with
$N_e=2$ and $10$, and all authors. Authors grew 10.6\% per annum at
all experience levels, correlating with post-PhD participation in research publications.

Growth stems from increasing enrollments in research programs and yet there is still unmet demand for computer science graduates and computer science innovations. This growth is critical for continued progress in computing. Publication practices will need to innovate to keep this growth from overwhelming the research publication process while maintaining rigorous peer reviewing standards, giving high quality feedback to authors, and moderating reviewer workloads. For example, communities are experimenting with practices such as tiered reviewing, increasing participation in the reviewing process, and limiting repeated reviewing of rejected submissions to multiple venues.  We need these innovations and more to support individual researcher’s careers and scientific progress.


\textbf{(2) Research quantity and quality are not correlated.} As is
well recognized, the volume of research output is not and should not be
used as a proxy for research
quality~\cite{CAL:2015,CRA:TPH:2015,JudgingQuality:2003}, but nor is
high output an indicator of low quality or vice versa. Research output
per author has actually declined between 1990 and 2012 as measured by
fractional authorship. Figure~\ref{fig:acm-fraction-papers-per-author}
shows that the weighted publications per author. Each author on
a paper accrues a fractional publication as a function of the total
number of authors. We plot the 50th (median), 90th, and 99th percentile
authors.  Figure~\ref{fig:acm-raw-papers-per-author} shows while the raw
number of papers for the 99th most prolific authors has grown
steadily,  the 90th percentile of authors produce one or two papers
per year. 
The most prolific 1\% of authors produced just two papers per year
when weighted for authorship contribution.  Unweighted, this
translates to five or more publications, whereas the 90th percentile
of authors produce one or two publications per year.
 Furthermore graduate student growth continues to outpace faculty
 growth~\cite{Taulbee:2015}. Because producing successful graduate students
 requires publishing with them, each faculty member should be publishing more, but they are not.
If some perceive that research quality has dropped over time, it is
\emph{not} due to an increase in per-author output.


The top conferences and institutions set research standards explicitly
and implicitly, in part because the U.S. computer science
professoriate is disproportionately populated by PhDs trained by four
U.S. departments, i.e., Berkeley, CMU, MIT, and
Stanford~\cite{CAL:2015, elitism:2015, Taulbee:2015}. We find that
authors in these four departments are equally prolific as compared to
all authors, measured both in raw publications and weighted by
fractional authorship (Figure~\ref{fig:acm-institutions}). We then restrict the comparison to established researchers, where an established researcher at N = 5, means their publications spanned five or more years. Established researchers at these four departments are slightly more prolific than all established researchers — suggesting that high output alone is not an indicator of poor quality research.


\textbf{(3)  Collaboration increases the impact of individual researchers.}
We find collaboration is highly correlated with quality.  In
particular, researchers at the top four institutions collaborate more
and papers in top conferences have more authors than other venues in
the same subdiscipline. Figure~\ref{fig:acm-percentile-avg-authors} shows shows the average number of authors
per paper has systematically grown and researchers in the top
departments collaborate significantly more than all authors on average
(p-value$<<$ 0.0001).  Furthermore, we find that higher rates of
collaboration are strongly correlated with quality. We analyze in more
detail the ten largest ACM Special Interest Groups (SIGs):  SIGCHI,
SIGGRAPH, SIGWEB, SIGDA, SIGIR, SIGSOFT, SIGARCH, SIGMOD, SIGPLAN, and
SIGMOBILE. We compare the top venue in each subdiscipline (SIG) to
collaboration practices for all venues in the SIG.  We find more average
authors on top venue papers with p-value $<$ 0.07 for 8 of 10 SIGs.
Only two (SIGIR and SIGARCH) were not significantly higher.
Furthermore, we add to the evidence that it is unsound to compare
publication practices (e.g., number of publications, citations, and
collaborations) of computer scientists in different
subdisciplines. For instance in 2012, SIGPLAN published 739 papers
with 3.13 average authors per paper, whereas SIGCHI published the most papers at
2562 in 2012 and SIGARCH had the highest average authors at 4.00.

Collaboration complicates the job of hiring and promotion committees,
which must evaluate individuals. Since collaboration correlates with
quality and collaboration is increasing over time, these committees
should expect successful researchers to exhibit a combination of
leadership, shared team leadership, and supporting roles in impactful
collaborative research. Instead of giving no credit to collaborative
research or forcing researchers to divy up credit, committees should
consider magnifying the credit to individuals in highly impactful
collaborations.  Furthermore collaboration is good for training
graduate students. They benefit from exposure to more research
ideas, research styles, and research area expertise. The whole is sometimes more than the sum of the parts.

We were surprised by several of these findings and it would be interesting to see what has happened since 2012. (We were delayed in disseminating this analysis since several fine venues chose not to publish this paper.) We sincerely thank ACM for providing a complete snapshot of their digital library.  We hope these findings spark some new conversations around growth and the value of collaboration.

Our summary
analysis is available as supplementary material on arXiv~\cite{BMX:ACM:analysis:2019}.

\section{Motivation: Publication Overload}

Science makes progress by identifying important problems, developing
new approaches, and incrementally improving existing approaches.
To handle growth, the computing
community has introduced more venues, more papers at the same venues,
tiered peer review, year-round conference submissions, and other
innovations.  However, the community is still under stress. Many
researchers have pointed out problems stemming from growth and
suggested improvements to the peer reviewing process in a spate of
CACM editorials, workshops, and suggestions for best practices~\cite{culture:2015,web:kathryn:2015,usenix-atc}.  This paper seeks to
understand better the source of growth to inform how the community
responds to this growth.

Analysis of science as a whole finds that there is exponential growth
in authors, publications, venues, and citations over
time~\cite{exponential:2000}.  These results indicate that as a
science matures, it diversifies and specializes, adding more
participants and their publications, and science progresses faster.
In this paper, we examine computer science trends using computer science author and publication
practices from the ACM Digital Library (DL), a large sample of
computer science publications, focusing on 1990 to 2012.

For our investigation, we mine ACM DL data on authors, institutions,
and venues for papers in ACM proceedings (conferences, symposia, and
workshops)~\cite{ACM:DL}. We examine subdiscipline trends based on
Special Interest Group (SIG) sponsorship and compare with prior
subdiscipline studies~\cite{Balzarotti:15,SE:Healthy:2014}. We
correlate author trends with faculty and PhD student growth from the
Taulbee Surveys of North American PhD-granting
institutions~\cite{Taulbee:2015}.  We consider research quantity and quality by
examining ten top conferences from the most prolific SIGs and
publication practices at Berkeley,
CMU, MIT, and Stanford. We choose these institutions because they are
well regarded 
and
their PhD students dominate Northern American
academic positions in computer science~\cite{CAL:2015,Taulbee:2015,elitism:2015}.

We address
questions such as: How is the field growing?  Have authors changed
their behaviors?  Is there a relationship between the quantity and quality of an individual's
research?  
Our subdiscipline
breakdown by Special Interest Groups (SIGs) is motivated by citation
analysis results that indicate citation practices should not be
compared across subdisciplines~\cite{exponential:2000}.  Our deeper
analysis of the citation patterns of ten top conferences exposes further
differences in subdiscipline citations practices over time and interactions between fields. We limit
our analysis to ten conferences because cleaning the data is time
consuming. 

\subsection{Overview of Findings}

This section overviews all our findings on publications, authors, venues, top
conferences, top departments, quantity, and
quality. Section~\ref{sec:methodology} presents our methodology for
using the ACM DL. Section~\ref{sec:analysis} presents our analysis of
authors and publications in detail. Section~\ref{sec:quality} examines quantity
and quality based on venues, discipline, and departments.

\paragraph*{Publications, authors, and venues}
We compare growth in publications (Figure~\ref{fig:acm-pubs}) and
authors (Figures~\ref{fig:acm-unique},
~\ref{fig:acm-papers-with-n-authors},
and~\ref{fig:acm-avg-authors}). We distinguish
established authors with publications spanning $N_e$ or more years with
$N_e=2, 5$, and $10$ and focus
on growth from 1990 to 2012. These results show that the
yearly increases in publications of 9.3\% per annum are explained by
yearly growth in the number of unique authors of 10.6\% per annum.
Figure~\ref{fig:acm-fraction-papers-per-author} shows that authors are
actually publishing \emph{less} when measured in fractional contribution to
each paper.
Collaboration has increased steadily and thus indvidual
authors are publishing slightly more when measured in raw papers
(Figure~\ref{fig:acm-raw-papers-per-author}).  In 2012, the most prolific 1\% of
authors produced five or more papers per year or 1.8 papers per year
weighted by co-authorship, accounting for 3\% of all publications.

\begin{quote}
  \emph{Exponential growth in the total number
of authors and more collaboration explains publication growth.}
\end{quote}

Individual authors are not publishing more than authors published in
the past. Therefore, \emph{if} the quality of publications is or has
changed, the explanation is not simply rooted in author output.




We next examine authors and venues in more detail.  Computing research
is increasingly a growing international
community~\cite{international:2012}.  The number of institutions
producing publications increased at \checked{10.8\%} per annum from
1990 to 2012, doubling every seven years, with most of this growth
from industry and non-U.S. institutions. While the authors from four
top U.S. departments continue to produce a disproportionate number of
publications compared to their numbers, their overall fraction is
dropping.
The growth in North American PhD graduates reported in the Taulbee
Survey~\cite{Taulbee:2015} correlates with publication
growth. The per annum growth rate of PhDs granted by North American
institutions from 2003 to 2014 was \checked{7.0\%}. A consequent
growth in publications is consistent with community standards that
require that PhD graduate students publish in order to graduate.

To understand venues and quality, we examined ten conferences
that are recognized as extremely high quality. We use sponsoring ACM Special
Interest Groups (SIGs) to represent subdisciplines and chose one high
quality (top) conference from each of the ten most active SIGs based
on the subdisciplines' judgements.  \checked{Any quality changes in these
venues over the period of our study are orthogonal to our analysis.}
The median growth in publications was 4.1\% per annum and the
acceptance rates at these conferences did not change.

\begin{quote}
  \emph{While ten of the ACM's top venues are publishing an increasingly small
    fraction of all publications, new venues are expanding research
    topics and scope.}
\end{quote}

Most growth in the ACM DL was instead due to new subdisciplines and
venues.  Whereas the number of ACM-sponsored conferences grew by 6.8\%
per annum, doubling every decade, the number of workshops grew far
faster, at 21.3\% per annum, doubling every three and a half years.  A
cursory examination of workshops shows they serve a wide variety of
functions. For example, they (a) jump-start topics that evolve into
conferences, (b) become part of existing conferences, (c) last a short
while, and (d) last for decades. New areas and deeper treatment of
some topics are still flourishing and likely to generate future
conferences and SIGs. Subdisciplines are experiencing different growth
rates. For instance, publications in SIGCHI, SIGSOFT,
SIGIR, and SIGCOMM grew by 11\% to 13\% per annum, substantially
faster than average.  Some new venues in these SIGs stem from
specialization. 

More often however, researchers established new SIGs,
such as SIGMOBILE, SIGBED, SIGKDD, SIGMM, SIGSAC, SIGWEB, and SIGHPC,
and venues that both
deepened and broadened research communties. The SIG's
statements of purpose, histories, and conference call for papers
reveal these trends.  For example, several new SIGs span
all aspects of their topics, e.g., the theory, programming systems,
runtimes, architectures, and applications of web, mobile, and embedded
systems. Many of the newest SIGs expanded
the fastest. 



\paragraph*{Quality}
We study the relationship between \emph{quality}, \emph{quantity},
and \emph{collaboration} by comparing the ten top venues to their
subdisciplines and four top U.S. computer science research
departments  (Berkeley, CMU, MIT and Stanford) to the field as a whole.  We note that quantifying
research publication quality and changes over time across all of
computer science is essentially impossible. Even expert
  reviewers have trouble judging the quality of a given publication. While detecting poor quality
  research is relatively easily, examining best paper awards
  retrospectively shows that relative judgements on high quality work
  have poorer predictive power than author
  productivity~\cite{JudgingQuality:2003}.

We find that the top conferences sustained the \emph{same} paper acceptance rates,
while the number of submissions and accepted papers grew at a median
rate of 4.1\% per
annum. 
The publications in the respective parent SIGs grew faster, at a
median rate of 10\% per annum, about double the rate of these top tier
conferences, reflecting a growing number of venues in each SIG.  The
number of authors on publications at top venues grew fifty
percent faster than publications because authors collaborated
more. Compared to the ACM and the sponsoring SIGs, which represent the
distinct subdisciplines, the papers at top conferences have more
authors. 


\begin{quote}
  \emph{Publications in eight of ten top conferences have
    statistically significantly more authors on average than
   than their subdiscipline  and than the ACM as a whole.}
\end{quote}

One quality judgement the academic community makes is through faculty
hiring~\cite{CAL:2015}.  Curated Taulbee
data~\cite{Taulbee:2015,elitism:2015} shows that PhDs from four
top departments disproportionately dominate the professoriate at the
top 25 institutions (43 to 59\% of all faculty in 2014).
Figure~\ref{fig:acm-avg-authors} shows that four top U.S. computer
science departments collaborate more than the rest of the field,
averaging 3.9 authors per paper in 2012, compared to 3.4 for the field
on the whole.  However, their output is the same as the field,
with the exception of their established and most prolific 1\% authors,
which publish slightly more than everyone else.  The 90th percentile
of researchers at four top departments publish the same modest number of
papers (one or two) per year as the rest of the field, although the
established 90th percentile at the top four publish slightly more
(four or more papers) than other established authors in the field.



\begin{quote}
  \emph{Compared to all authors, researchers at four top
    U.S. departments are similarly prolific, but collaborate more. The
    most prolific established researchers at the top four
    departments publish slightly more than other established
    researchers.}
\end{quote}


\section{Methodology}
\label{sec:methods}
\label{sec:methodology}

This section presents our methodology in detail, describing the ACM
Digital Library data we use and our analysis process.

We analyze publications and authors in the ACM Digital Library (DL)
corpus~\cite{ACM:DL}, which incudes every article from conferences, symposium,
journals, magazines, workshops, etc. that the ACM published since
1954.  We present data from 1954, but focus on recent growth from 1990
to 2012, the last year of complete data when we started this project. 
The ACM graciously provided us with a download of the
entire database of articles in late 2014 which includes metadata such
as author names, venue, and date of
publication.
We restrict our analysis to ACM sponsored proceedings which include conferences, symposia, and workshops. The
full list of venues and publications are available from the DL
\url{http://librarians.acm.org/digital-library}.  We classify
proceedings by the sponsoring SIG, using DL metadata, and use them to
explore subdisciplines. Co-sponsored events accrue to all sponsoring
SIGs in the individual SIG analysis, but only once for ACM
totals. This corpus represents a large and representative fraction of
computing publications, but is not exhaustive. For example, some
USENIX conferences, several top security venues, IEEE venues, and many international
computer science venues are not included. Many of these publications
are in the DL.
Because, however, only ACM-sponsored venues are systematically included
in the corpus, we limit our analysis to ACM-sponsored venues. These
venues do include co-sponsored venues, e.g., co-sponsored with IEEE and others.
We limit our top conference analysis to ten because we found classification
errors on publication types (see below) that
required that we hand-correct and hand-verify the analysis; a time consuming process.

Large data sets such as this one inevitably contain errors.  During
verification of the ten sampled conferences, we found a number of
classification errors, which we fixed in our analysis and reported to the
ACM. For
instance, many of the conferences including CHI, OOPSLA, and SIGRAPH
include posters and/or workshop publications in the main conference
proceedings and either do not use the metadata to label them, or
incorrectly label them as conference papers.  To make things worse,
these errors are not systematic across instances of a given
conference.  Because this type of validation is time consuming, we did not analyze more
individual conferences.  We urge each SIG to devote resources to
checking and correcting their metadata, venues names, and addressing other
errors.  

We use the ACM's
author identifiers for the analysis, which contain errors due
to name aliasing, authors who change their name, and authors who spell
their names
inconsistently, such as different initials or
middle name usage.

To understand author trends over time in more detail, we define an
\emph{established author} as someone (with the same author identifier)
who publishes papers spanning $N$ calendar years, with $N_e$ = 1 to
10.  When $N_e$ = 1, all authors are called `established'.
When $N_e$ = 2, authors become established in their second year; when
$N_e$ = 5, authors are established in the fifth year, i.e., four years
after their first publication.    We use $N_e$ = 5 by default, but found
that our analysis was not very sensitive to the choice of $N_e$~\cite{BMX:ACM:analysis:2019}. When restricting analysis to established authors, we
remove the impact of authors who only publish once and many students who
do not continue to publish after completing their degree. Analysis of
established authors focuses on career behavior in academia, industry,
and government.

We report publishing institutions per year, which has a systematic over-reporting
bias. On occasion author affiliation is missing.  More
common is that institutions have multiple names, over-reporting the
total number of institutions. Author reporting of institution is more
inconsistent than names.  Institutions vary in the policy that the ACM
enforces.  We analyzed data for four top U.S. departments, checking for aliases by hand. As one might expect, ACM captures Stanford publications as coming from the corresponding single
institution.  On the other hand, MIT, CMU, and Berkeley have multiple
department and lab institution identifiers in the DL. The DL counts each of these separately. Ambiguity also
stems from Universities with multiple campuses. These errors tend to
over-report the number of institutions.

To analyze author growth in more detail, we use some Taulbee survey
data~\cite{Taulbee:2015}. Since 1971 this survey has gathered
enrollment, production, and employment information of PhDs and faculty
in North American (U.S. and Canadian) computer science (CS) and
computer engineering (CE) departments, and recently added information
systems (IS) departments.  Similar data is unfortunately not available
from other countries or regions.  The Taulbee survey held department
rank constant for the top 104 in this period for analysis purposes. We
compute tenure-track faculty and PhD production per annum growth rates
using data from 2003 and 2014 from 104 reporting and ranked CS and CE
departments obtained from Betsy Bizot at the Computing Research
Association (CRA) --- the data for these years and departments was
easily available.

To examine quality, we consider the behavior of authors at four top
departments: Berkeley, CMU, MIT, and Stanford.  Taulbee ranks these
four departments as the top departments.  One distributed judgement
of quality is faculty hiring, which supports this
ranking.  Clauset et al.\/ show that the hiring practices over 205
North American computer science departments between May 2011 and
March 2012, suggests a ranking of Stanford (1), Berkeley (2), and MIT
(3), CMU (7), and Cal Tech (4)~\cite{CAL:2015}. (The size of PhD
granting department is not a statistically significant factor in PhD
placement~\cite{CAL:2015}.)  We use the 2003 and 2014 Taulbee data to
report current PhD employment as a function of PhD granting
institution in the top ranked departments~\cite{elitism:2015} and then
compare publication practices at the top four to the wider
community. This ranking selects the top four we use in our study. 

%
%

\section{Publication Trends}

This section starts with aggregate author and publication trends
over time, and then examines departments, institutions, venues, and
subdisciplines in more detail.

\subsection{Publications and Authors}
\label{sec:analysis}
Figure~\ref{fig:acm-pubs} presents the articles newly published in
proceedings each year. In the modern computing era between 1990 to
2012, the number of published articles per year grew exponentially
from 2,650 to 14,521, a factor of 5.5 (\checked{9.3\%}  per annum) --- not quite as fast as the doubling of transistors every
two or so years delivered by hardware manufacturers. This substantial
growth in publications is hardly a surprise.

%
%

In the same period, the number of unique authors per year grew
\emph{faster}, at \checked{10.6}\% per annum 
from 4,865 authors to 
35,725, a factor of 7.3. Figure~\ref{fig:acm-unique} plots the number
of unique authors and established authors by year. In 1990, there were
\checked{1253 unique established authors ($N_e$=5), which grew to 15,232
  in 2012, a factor of 12.2.}  
The growth in established authors is interesting when coupled with
with the slower rate of U.S. faculty growth (see
Section~\ref{sec:phd-faculty}) and recent findings that document
substantial growth in international computer science
research~\cite{international:2012}. Together they suggest more
research participation world-wide coming from industry and government,
not just academia.


The growth in active authors each year is higher than the growth in
publications, but this difference is only in part explained by increases in collaboration.
Figure~\ref{fig:acm-avg-authors} plots the average number of authors
per paper, which has risen from 2.1 in 1990 to 3.4 in 2012.   Since the mid-eighties, the average number of authors per paper from the top four departments has consistently been higher than for the ACM as a whole.
Figure~\ref{fig:acm-papers-with-n-authors} plots a line as a function
of year and shades the region for the percentage of publications with
one, two, three, four, and five or more authors. The lowest line is
one author ($N = 1$).  Since the late 1960s, publications increasingly have more
authors every year, but as late as 1980, 50\% of publications had only
one author.  In 2012, 23\% of publications have four or more
authors, whereas in 1990 just 6\% did. The proportion of single author
papers per year has declined over time from 34\% in 1990 to 11\% in
2012. Section~\ref{sec:quality} shows that ten top conferences and
researchers from four top U.S. schools have
even more
authors on average.









Several hypotheses may explain the increase in collaborations. For
example, a cultural change may have occurred in which advisors play a
larger role in the research process, or their role is acknowledged
more, and thus faculty appear on more of their students
publications. Increases in collaboration may be due to the type of
research.  Computer science is increasingly impinging upon and enabling other disciplines.  Multidisciplinary research that combines distinct
sub-disciplines in computing or other fields, such as computational
biology or finance requires more expertise, which is easier to acquire by adding
a co-author compared to earning a degree or otherwise gaining
sufficient background in another field.  Larger more sophisticated and
ambitious projects may require more people to build and understand the entire
system. For instance, when the DaCapo NSF ITR project built
performance evaluation methodologies for managed languages and a
Java benchmark suite~\cite{DaCapo}, it required developing new tools; new
analysis; new data sets; modifying about 30 active open-source
projects; and then evaluating and discarding some candidate benchmarks. The
result was a publication with 20 authors. The data shows
collaboration is increasingly common.

\begin{figure}

\includegraphics[width=.9\columnwidth]{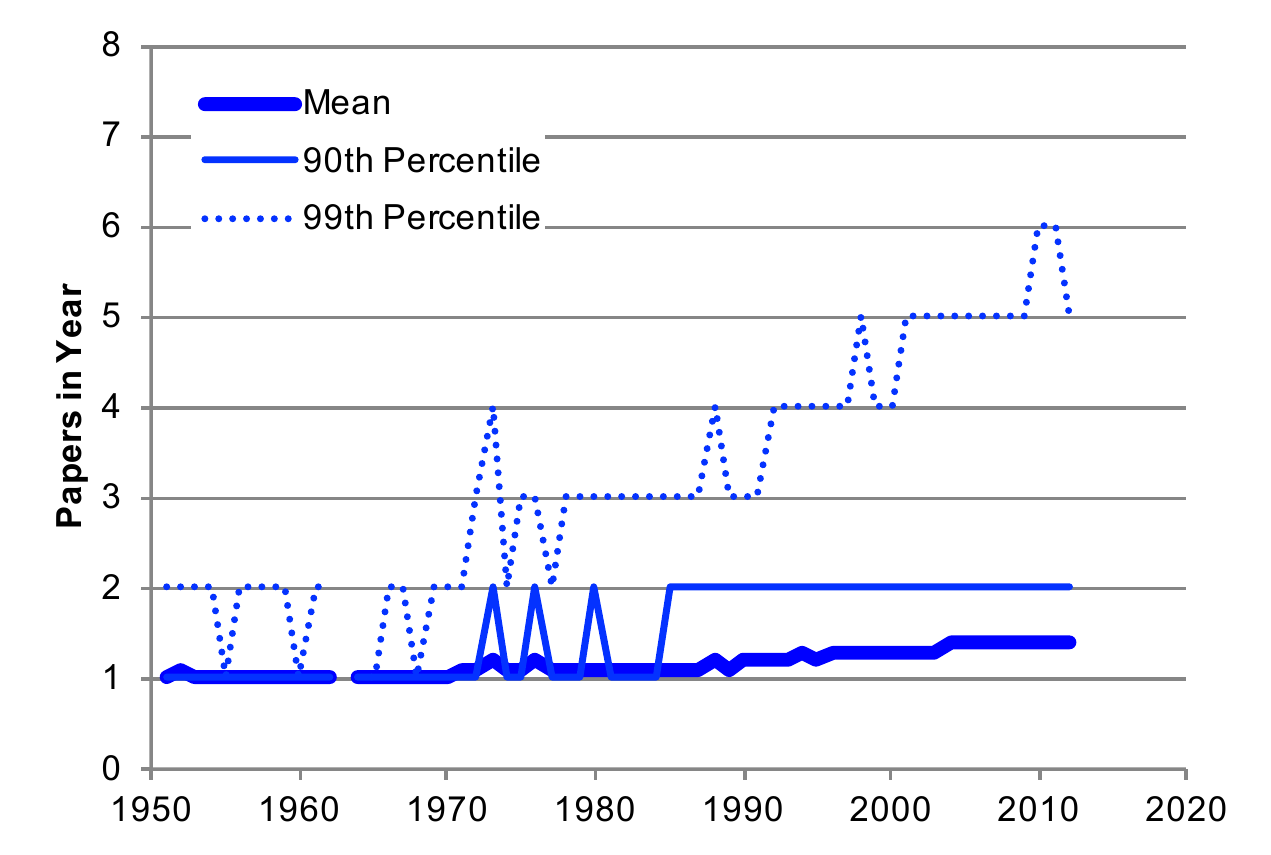}

\caption{Raw publications per year per author.  We report the mean for all authors and the raw publications for the 90th and 99th percentile authors.}
\label{fig:acm-raw-papers-per-author}

\includegraphics[width=.9\columnwidth]{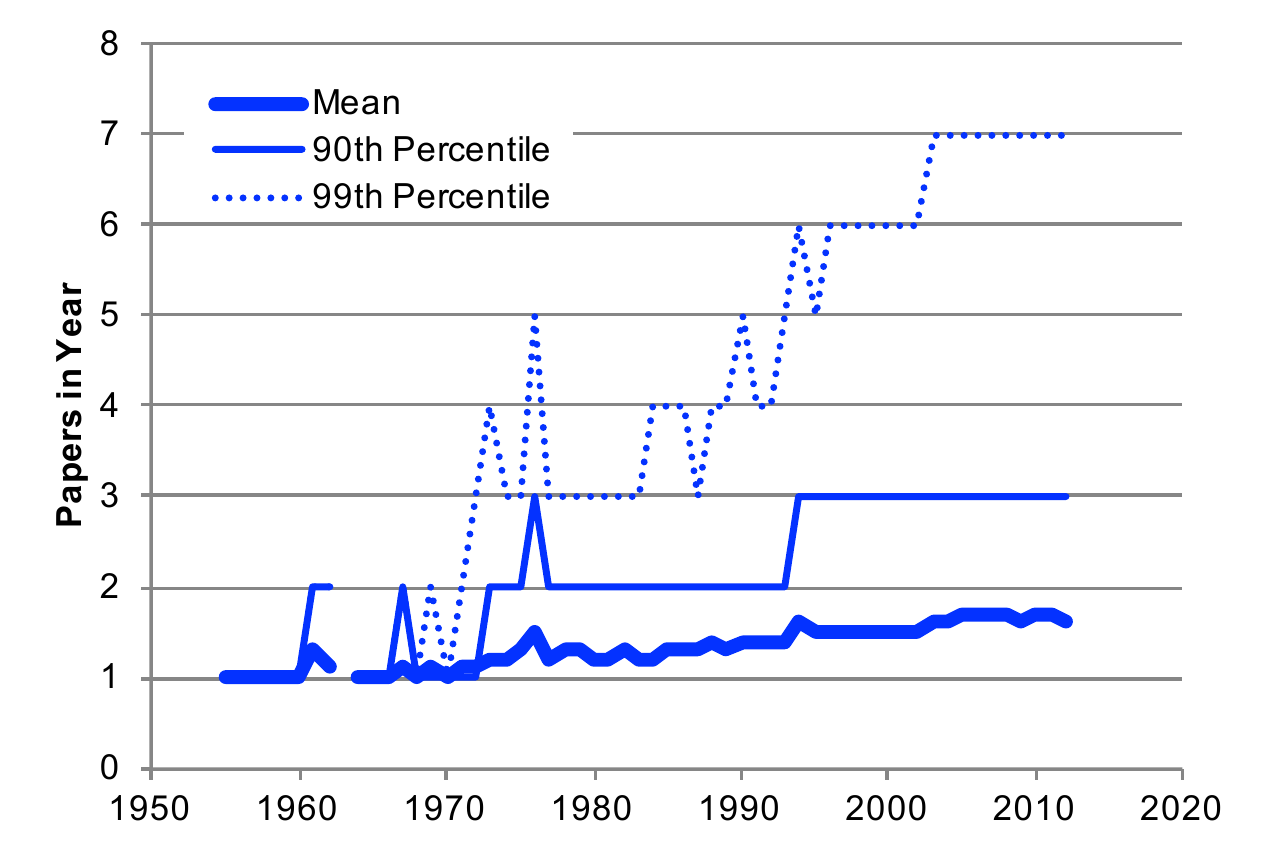}

\caption{Raw publications per \emph{established} author ($N_e=5$).}
\label{fig:acm-raw-papers-per-established-author}


\includegraphics[width=.9\columnwidth]{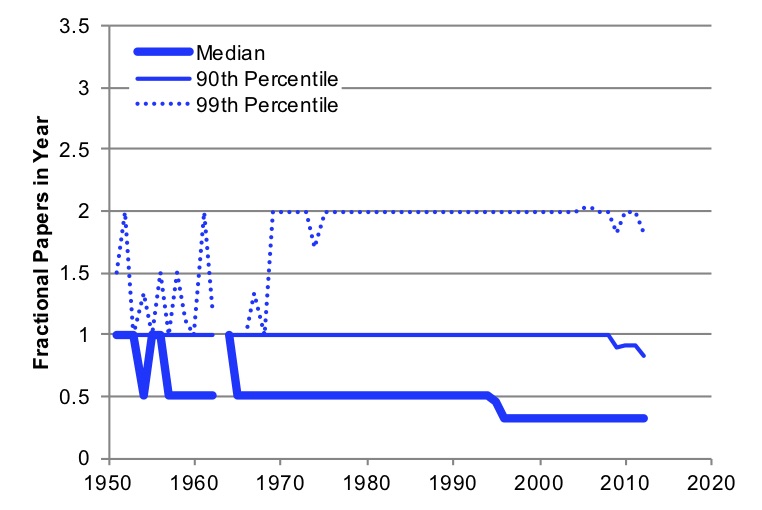}

\caption{Weighted publications per year per author. We report the weighted number of publications for the 50th (median), 90th, and 99th percentile authors.}
\label{fig:acm-fraction-papers-per-author}

\includegraphics[width=.9\columnwidth]{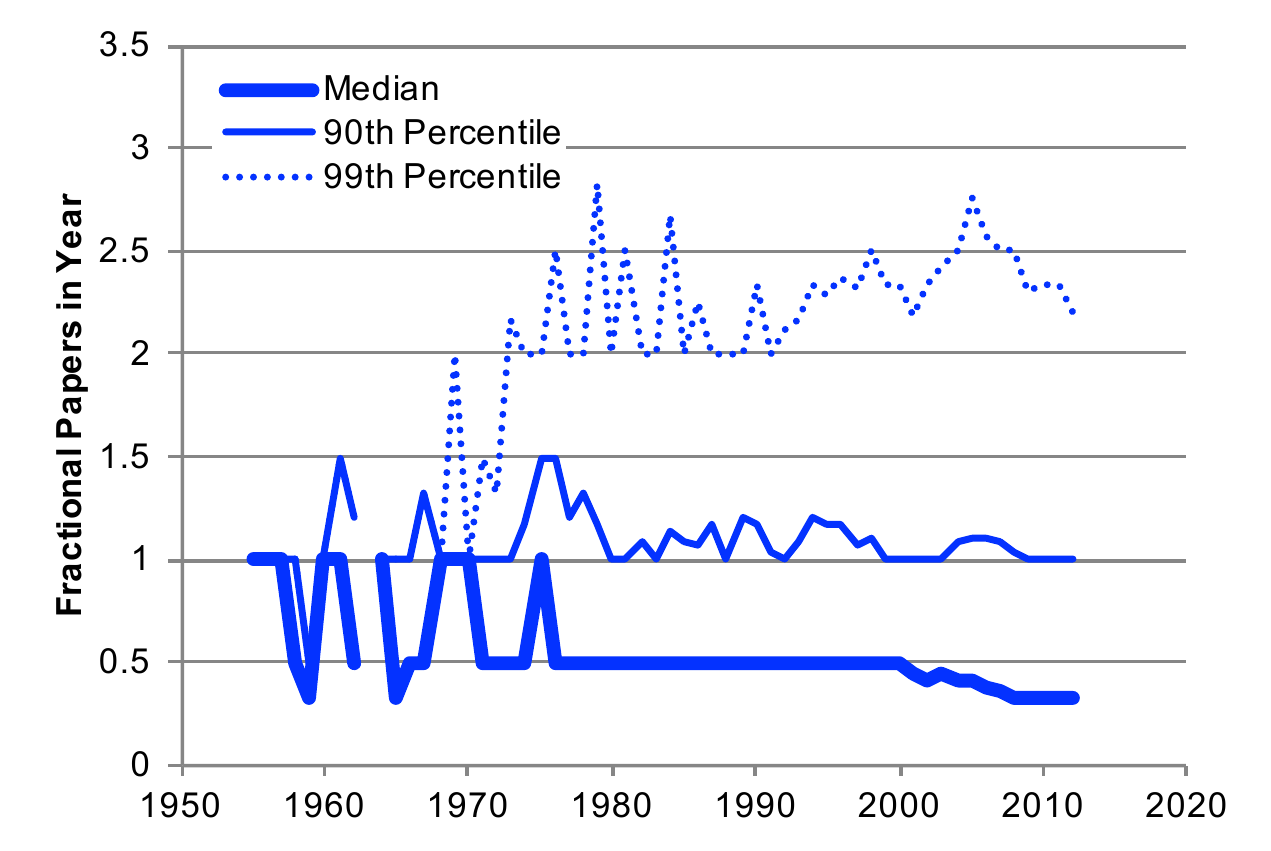}

\caption{Weighted publications per \emph{established} author ($N_e=5$)}
\label{fig:acm-fraction-papers-per-established-author}

\end{figure}

Figures~\ref{fig:acm-raw-papers-per-author}
and~\ref{fig:acm-raw-papers-per-established-author} show the
\emph{raw} mean number of publications per author and per
established author per year, as well as the number of publications per author for the 90th and 99th percentiles.  Note that these statistics are \emph{not} computed per
individual author year over year. The mean publications per year
per author rose from 1.2 in 1990 to 1.4 in 2012, 0.7\%  per annum.\footnote{We report the mean rather than median since publications per author yeild a small integer value with a median value that is invariably 1.} The 90th and
99th percentiles report number of papers published by the author at the respective percentile.  For the most prolific 1\% of authors, raw papers per year has
risen from 3 and 4 to 6.  Among established authors, the mean
publications per year per author grew slowly from \checked{1.4 to
1.7. The 99th percentile most prolific established authors had 5
publications per year in 1990 and now have 7.} 
These publications comprised 1,412 out of 14,521 papers in 2012 (9.7\%).
Consequently, although eye-catching, these authors have relatively low impact on the field as
a whole. Whereas 90\% of all authors produce one or two papers each
year, 90\% of established authors produce one, two, or three
papers.

Figures~\ref{fig:acm-fraction-papers-per-author} and~\ref{fig:acm-fraction-papers-per-established-author} show the
\emph{fraction} of publications per author and per established author
per year, for the median as well as the 90th percentile and 99th percentiles. We compute the
fractional publications per author by partitioning the publication
equally among the authors and summing for each author each
year. Author output has declined since 1990 by this metric. The median
author was responsible for 0.5 publications in 1990, and in 2012 is
responsible for 0.33 publications.\footnote{We report median rather than mean because this indicates the behavior of the `typical' author behaves, whereas the mean is skewed from the median by exceptional few who publish prodigiously.} The 90th percentile of all and
established authors follows the same trend. The 99th percentile authors
dropped from 2 publications to 1.8. The median established author
has the same trend, dropping from 0.5 to 0.33 publications per
year. The 99th percentile fluctuates more in this period: 2.33 in
2000, to a high of 2.81 in 2005, and a low of 2.25 in 2012. This data
taken together with the rising average number of authors per
publication shows that author output is relatively constant,
even for the most prolific authors, when normalized by author count.

 \begin{table}

\begin{tabular}{@{}r|@{}rrrrr@{$\;$}r@{}}
\multicolumn{1}{c}{\textsf{\textbf{Rank of}} } &
\multicolumn{3}{c}{\textsf{\textbf{Tenure Track Faculty}}} &
\multicolumn{3}{c}{\textsf{\textbf{PhD Production}}} \\
\multicolumn{1}{@{}c}{\textsf{\textbf{Institution}}} &
\textsf{\textbf{2003}} & \textsf{\textbf{2014}} & \textsf{\textbf{change}} &
\textsf{\textbf{2003}} & \textsf{\textbf{2014}} & \textsf{\textbf{change}} \\ \midrule
\textsf{\textbf{$\leq$ 4}} &  202 & 252 & 25\% & 80 & 161 & x 2.0 \\
\textsf{\textbf{$\leq$ 10}} & 445 & 547 &23\% & 162 & 328 & x 2.0 \\
\textsf{\textbf{$\leq$ 15}} &628 & 750 & 20\% & 230 & 457 & x 2.0 \\
\textsf{\textbf{$\leq$ 25}} & 910& 1039 & 14\% & 323 & 612 & x 1.9 \\
\textsf{\textbf{reporting 104}} & $\;$2714 & 3176 & 17\%& 670 & 1404 & x 2.1 \\
 \multicolumn{1}{r}{} & \multicolumn{3}{r}{\textsf{\textbf{per annum ~~~1.4\%}}}
& \multicolumn{3}{r}{\textsf{\textbf{per annum ~~~~~7.0\%}}}
\end{tabular}
\caption{\textsf{\textbf{North American Tenure Track Faculty and
      PhDs}} by institution rank  from the Taulbee Survey~\cite{elitism:2015}.
  While PhDs have doubled,  numbers of faculty grew more slowly. Four
  top ranked U.S. departments
  disproportionately produce PhDs.}
\label{fig:faculty-phd}
\end{table}

\subsection{North American PhD and faculty growth}
\label{sec:phd-faculty}

This section examines the correlation in author growth with PhD and
faculty growth.  Table~\ref{fig:faculty-phd} presents the growth in
North American PhD production per year between 2003
and 2014 and faculty size for 104 ranked departments from the
Taulbee data. PhD production increased at 7\% per annum,
whereas the number of faculty grew by 1.4\% per annum.  On average,
an increase in PhD students per faculty member represents an increase
in workload.
Comparing with author growth rate (10.6\% per annum) and
established researcher growth rate, this reveals that much author
growth is from authors in student cohort, industry, government, and international
institutions, rather than North American faculty~\cite{international:2012}.

Note that the faculty of top ranked departments grew more than lower
ranked departments. Furthermore, overall they produce a
disproportionate fraction of North American PhDs --- 44\% of PhDs
graduated from the top 25 ranked schools in 2014, slightly lower than
the 48\% in 2003. 
The fraction of PhD students from four top schools
has not changed --- four schools have disproportionally produced 12\%
of all U.S. PhDs since at least 2003. 

\begin{figure}
  \includegraphics[width=\columnwidth]{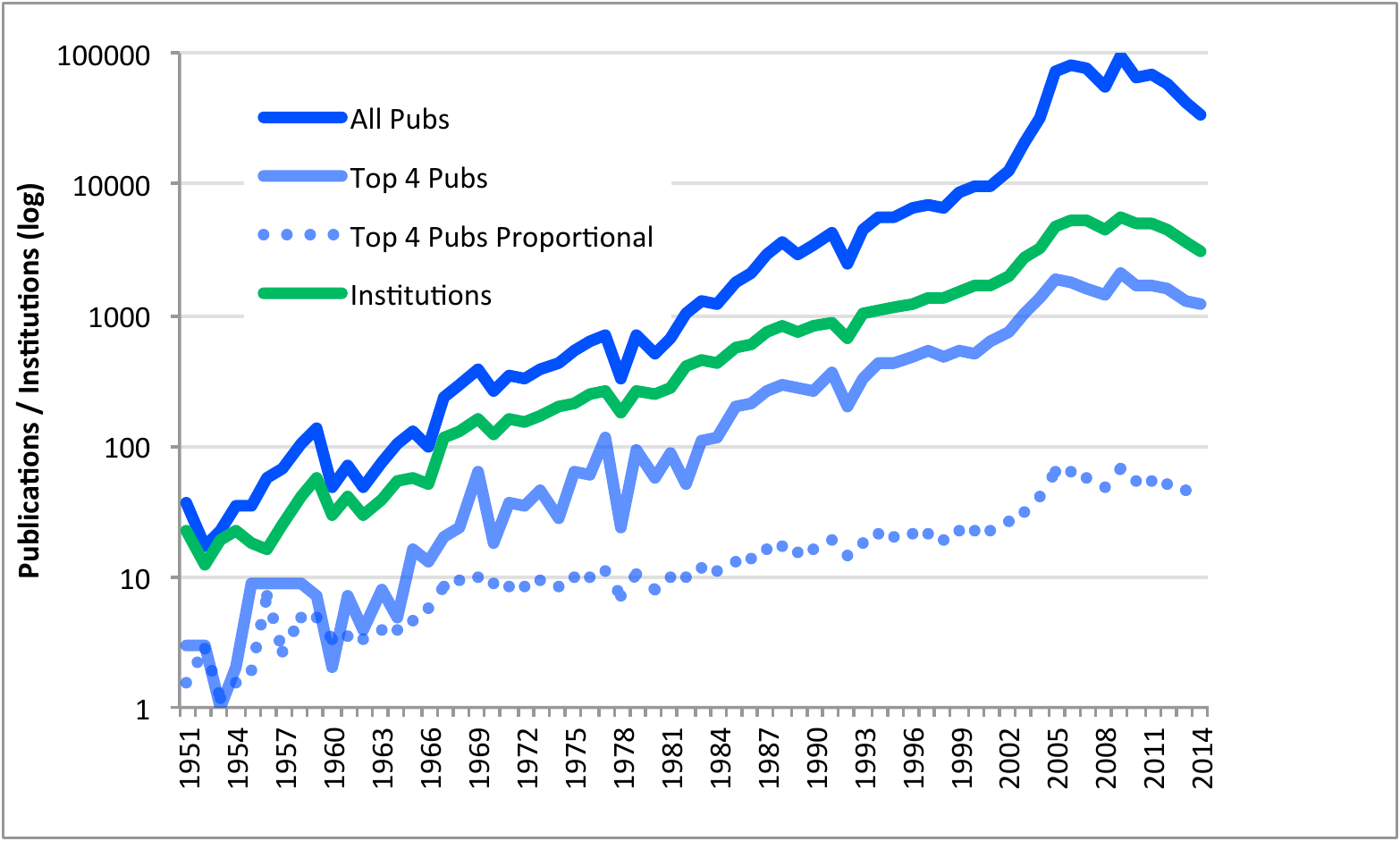}
  \caption{Unique institution growth measured by the ACM on every
    publication in their system, including journals, other publishers,
    etc. (All other graphs use ACM publications as described in
    Section~\ref{sec:methods}). Note the log scale. The fraction due
    to the top four is declining as the numbers of institutions
    participating in computer science research flourishes world wide.
    However, these four top departments still disproportionally
    produce publications --- the dashed line shows the number
    predicted by their institutional
    representation.} \label{fig:acm-institutions}
\end{figure}

\subsection{Institutions}

Figure~\ref{fig:acm-institutions} plots the number of unique
institutions with one or more publications each year and compares it
with growth in papers. We include each institution only once
regardless of the number of authors or papers. The ACM directly computed only
this data on all publications in the DL, including ones not published
by the ACM (without the restrictions described
in Section~\ref{sec:methods}).  Note that the publications are close to
100,000. We did not use this dataset for the other analysis because
many non-ACM venues were not consistently included, as shown by
the declines. The number of institutions grew at a
rapid pace, 8\% per annum from 1990 to 2012. Growth in institutions
would likely be lower if the institutional data reporting was more
systematic (see Section~\ref{sec:methods}). 
According to the 2014 Taulbee report~\cite{Taulbee:report:2014} (page
2), in 1995 there were 133 U.S. CS departments and in 2014, there were
188, a growth rate of \checked{1.8\%} per annum. Total CS, CE, and IS departments in the
U.S. and Canada grew slightly faster at a rate of
\checked{2.7\%} per annum. Institutional growth thus stems more from increases in
international academic participation in computing research, as well as
government and industry world-wide, as documented by previous
work~\cite{international:2012}.  Since participant growth outstrips
institutional growth, the unique researchers at other institutions
must be growing as
well.  


\subsection{Venues}

\begin{figure}
  \includegraphics[width=\columnwidth]{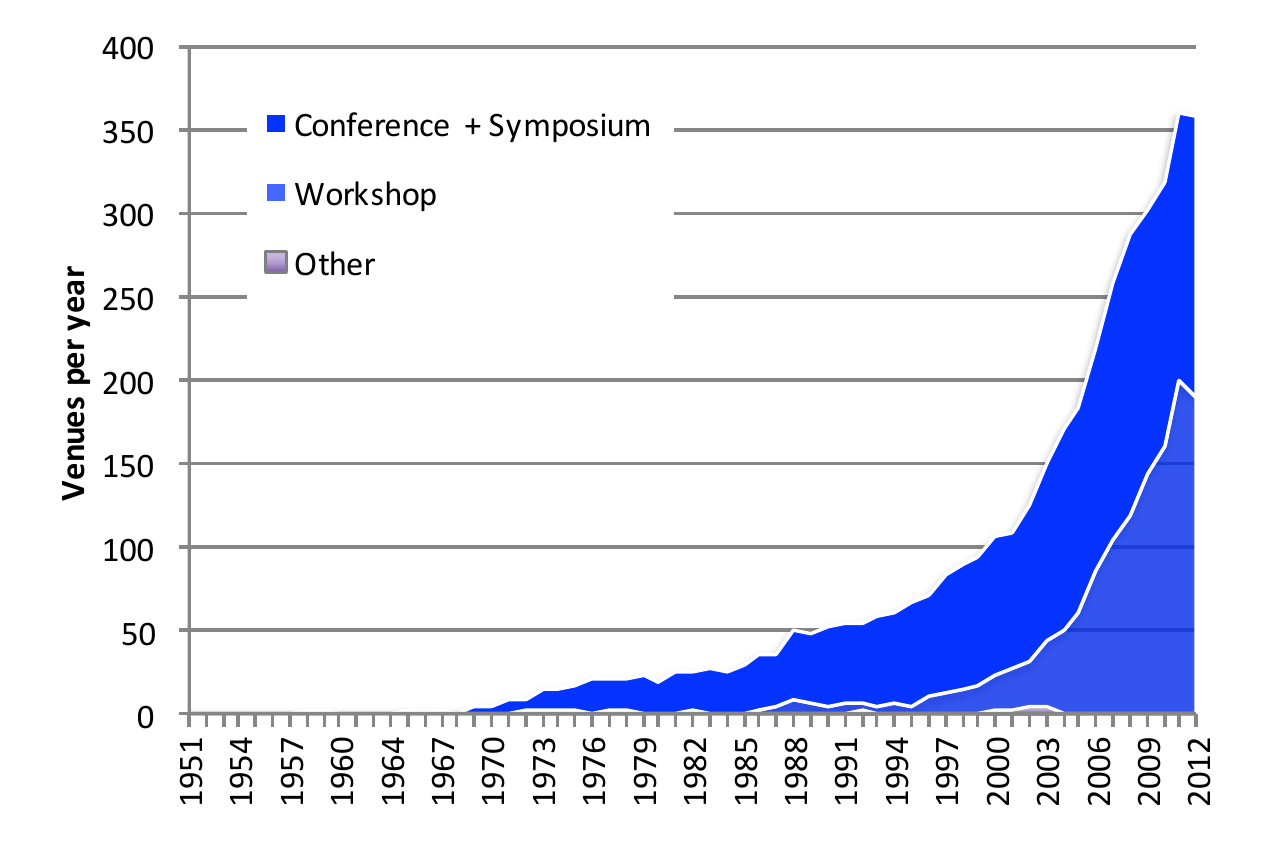}
\caption{Venues: Workshops are growing faster than conferences.}\label{fig:acm-venues}
\end{figure}

Figure~\ref{fig:acm-venues} presents the total number of conferences
(identified as `conference' and `symposium'), workshops, and other
proceedings publications published by the ACM. 
 The number of ACM-sponsored conference venues grew
at 5.9\% per annum from 47 to 167, but has been relatively flat
since 2007. On the other hand, the number of workshops
grew at 17\% per annum, from 6 to 190. If workshop organizers are
becoming more likely to include their proceedings in the DL, then the
growth rate for workshops would be over-reported.  The growth in
venues is slightly more than the growth in publications, but well below
the growth in authors. Workshops are a predictor of new
research directions that subsequently create SIG and conference
venues, and thus signal that the field continues to grow.

\subsection{Subdiscipline analysis by SIG}

This section examines growth by subdisciplines based on the sponsoring
ACM Special Interest Group (SIG). Table~\ref{tab:sigs} presents the
sponsoring SIG, the number of publications and average number of
authors in 2012. Columns four and five present growth rate per annum
for publications and authors from the period 1990 to 2012 (except
where noted in the last column).  We order the table by the number of
publications in 2012.  If a conference is co-sponsored by multiple
SIGs, we credit it to both SIGs in this analysis --- 3,102
publications fall into this category. The sum of all the SIG
publications (17,623) is thus larger than the last row, which presents
all proceedings papers in the ACM DL (14,521) from 1990 to 2012, where
each publication is counted only once.  
The table clearly shows the distinct collaboration patterns among
subdisciplines, with systems-oriented SIGS such as SIGCOMM (4.16),
SIGARCH, SIGOPS, SIGBE and SIGMOBILE having the highest average
authorship, and SIGACT (2.66) having the lowest.  
This finding is consistent with
statistically significant differences found between citation and
collaboration networks among disciplines and
subdisciplines~\cite{BMG:2004}.  In other words, it is
unsound to compare scientists in different disciplines and
subdisciplines based on collaboration, publication, and citation
patterns.

\begin{table}
\input{sig-table}
\caption{Publications, average authors, and per annum papers and author growth by subdiscipline
  based on sponsoring SIG from 1990 to 2012, except where
  noted, ordered by 2012. }
\label{tab:sigs}
\end{table}

SIGCHI has the most publications and
their number has grown at an above-average rate --- 15\% per annum
compared to 9.3\% of ACM as a whole. Eight SIGs have very high
growth rates:
SIGWEB, SIGIR, SIGMOBILE, SIGKDD, SEBED, SIGSA, SIGSPATIAL,
and SIGACCESS.  While the least
prolific SIGs generally have low growth rates, several of the newest
SIGs (SIGSPATIAL, SIGACCESS, and SIGecom) are not yet
prolific, but they are growing at above average rates (19, 15, and
10\% per annum respectively).

Subdiscipline growth in computing has spurred new venues and new
special interest groups.
The addition of fifteen SIGs and the demise of two in this period, as
noted in the table, shows both the dynamism of our field and the
growth in subdisciplines.
The new venues are not only
specializing. Many of the new SIGs seek to both deepen the treatment
of a topic and at the same time broaden it.
For instance, Victor Bahl and Imrich
Chlamtac, researchers who were publishing often in SIGCOMM and IEEE
INFOCOM, started SIGMOBILE in 1996 to deepen the treatment of
wireless networking as a distinct discipline and to establish a broader
community studying mobility of systems, users, data, and
computing~\cite{web:sigmobile}. SIGMOBILE help establish the community and clearly remains an important research venue. Publications in SIGMOBILE grew at the second most rapid
rate of all the SIGs at 26\% per annum and participants grew at 27\%
per annum. Only SIGBED grew faster, at
28\% (publications) and 29\% (authors) per annum.  SIGBED is a similar
example of a SIG that was created to cover all
aspects of a topic, in this case, embedded computing. Whereas SIGHPC
was formed to recognize an existing community,
most new SIGs, e.g., MOBISYS and SIGBED, built new communities.
These SIGs are breaking down
some of the traditional subdiscipline boundaries between theory,
compilers, hardware, networking, and applications.
New SIGs are defining communities with shared
    research interests, often beyond traditional subdiscipline
    boundaries.

\begin{table*}
\centering
\input{conference-table}
\caption{Per annum growth in papers and average authors for top
  conferences and their sponsoring SIG; 2012
  average authors, conference accept rates in 2012, and changes
  between 1990 and 2012.  ~$^{*}$2011 for SIGGRAPH since they became a
  journal in 2012.}
\label{tab:conferences}
\end{table*}

\subsection{Summary}  Most publication growth stems from new
conference and workshop venues that deepen and
diversify research areas.  Individual author output has
declined, even for the most prolific authors, when measured in
fractional authorship.  When measured in raw publications, authors
have modestly increased their output over this period, mostly by
collaborating more.  The very modest increases in raw output for
the average author, established authors, and even the most productive
authors (all less than 2\% per annum) is not the source of the 9.3\% per
annum growth in total publications.
The growth in computer science publications stems
    from the growth of the field as whole --- more researchers working
  on more topics.

\section{Quality}
\label{sec:quality}

This section examines quality from the perspective of venue and
researcher institution. (1) We sample ten top conference venues.
These ten conferences differ from the field as a whole: they are growing more
slowly and their average paper has more authors than their
subdiscipline, as represented by their sponsoring SIG.  (2) We examine
departmental hiring, PhD production, and the research output of four
top ranked departments. These departments hired more faculty and
disproportionately produced faculty for other departments.
Researchers at these four departments are similarly productive as other
researchers, but they collaborate more.

\subsection{Venue Quality:  Top conferences}


This section considers a sample of top ACM conferences and compares
their trends with those of their parent SIGs and ACM overall.  From
each of the ten most prolific SIGs, we chose one of the most
prestigious venues based the SIG web page,
citation counts, and our personal experience. These conferences are widely considered as very
high quality and their high citation rates show that they
significantly influence scientific progress.
Table~\ref{tab:conferences} presents for each conference: its publication
growth rate and that of its parent SIG; its author growth rate and that of its parent SIG; the average number of authors per paper in 2012 for it
and its parent SIG;  the student T-test $p$-value comparing the
average number of authors per paper in the SIG
(minus the conference) and the conference in 2012;  whether the submissions are
double-blind; the acceptance rate in 2012; and finally, the change in paper acceptance rate
for the conference over the period.

Publication growth rates at conferences, except for WSDM, are much
lower than their parent SIGs and the ACM as a whole, varying from
8.2\% per annum for CHI to 1.3\% per annum for ISCA, with most between 2\% and
6\% per annum, with a median of 4.1\%, which is substantially lower
than the median 10.4\% per annum growth rate for their parent SIGs and
9.3\% for the ACM as a whole.
Growth in total author numbers is also consistently lower in the
conferences (except WSDM) than their parent SIG and ACM as a
whole. The conference
median growth rate is 6.3\% per annum, just half that of the parent SIGS.
The SIG authorship pool grew at 12.9\%.  Similar to the ACM as a whole,
average authors at the top conferences grew substantially faster than
publications.

In all cases, the average number of authors per paper in the top
conference is larger than the average number of authors on all papers
sponsored by the SIG.  We use the Student T-test to determine whether
this difference is statistically significant. We compare the average
authors on all publications in the sponsoring SIG minus the
conference, with the conference.  The table presents the
$p$-values. For all conferences except for ISCA and SIGIR, the
$p$-value $< 0.07$, which means that the higher number of authors per
paper is strongly correlated with the highest quality venue.

Finally, conferences have
essentially not changed their acceptance rates.
These results are consistent with the subjective assessment that the
top venues in our field are maintaining very high standards whilst
seeing a modest growth in publications and a somewhat larger growth in
author participation.  

\section{Visualizing venue citations}

\begin{figure*}
  \includegraphics[width=0.90\textwidth]{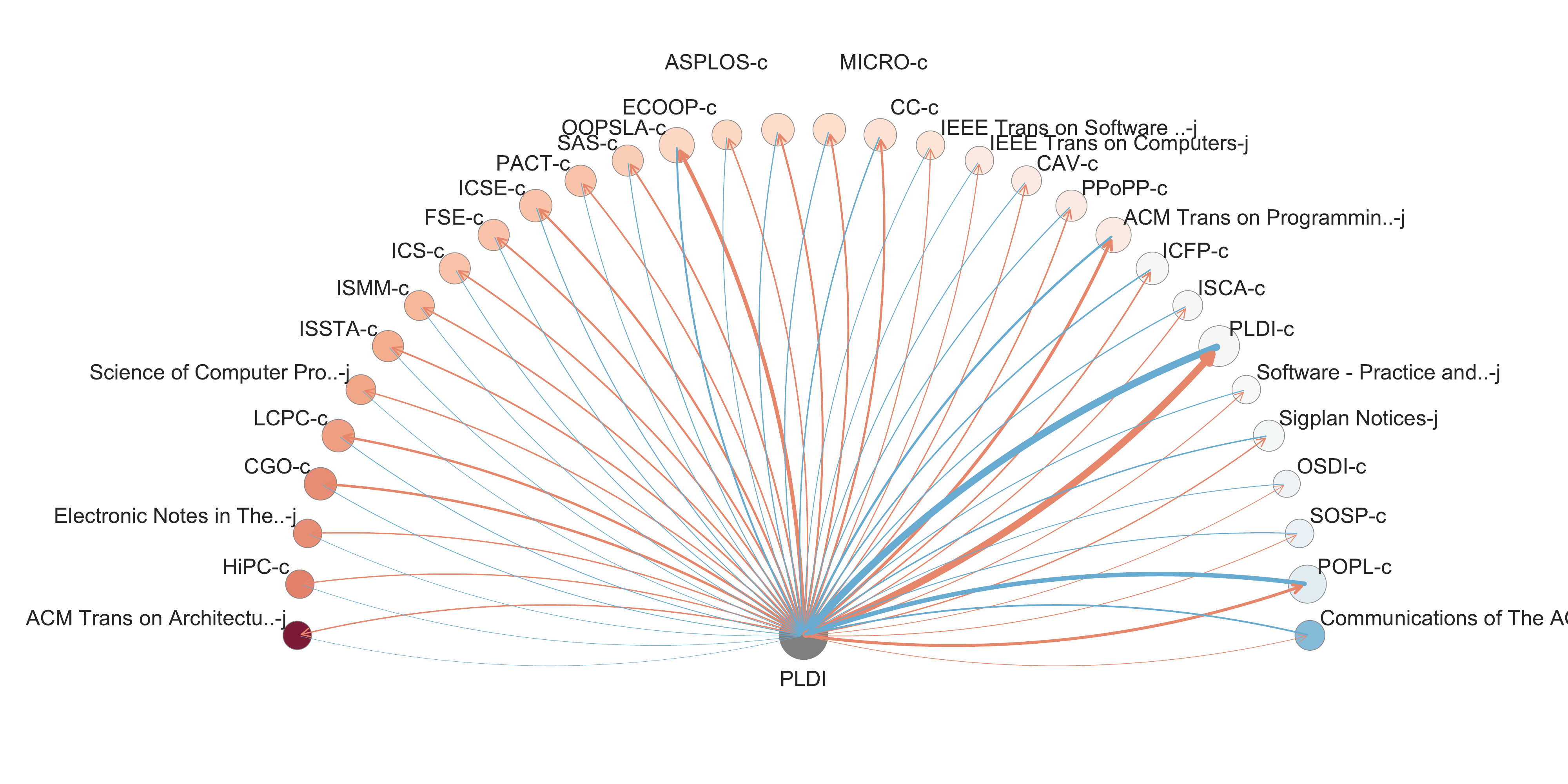}
  \caption{Summary of incoming versus outgoing citations to 25 top
    venues. Node colors: ratio of citations (outgoing ideas, red)
    versus references (incoming ideas, blue). Node sizes: amount of
    total citations (outgoing) and references (incoming) in either direction. Thickness of blue edges are scaled by the number of references going to a given venue; thickness of red edges are scaled by the number of citations coming from a given venue. Nodes are sorted left-to-right by the ratio of incoming vs outgoing citations to PLDI.}
  \label{fig:pldi-in-out}
\end{figure*}

\begin{figure}
  \includegraphics[width=0.35\textwidth]{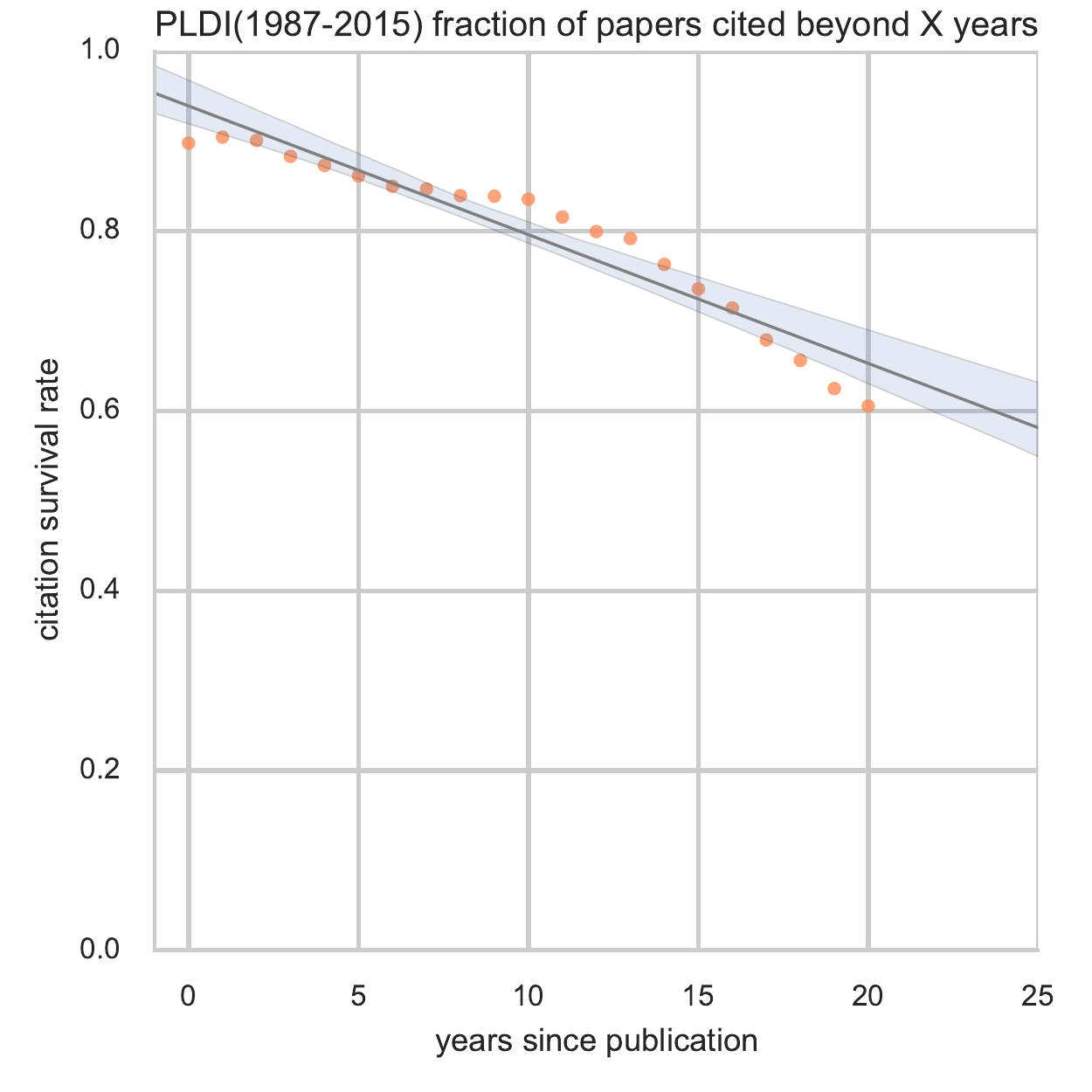}
\caption{Citation survival graph plots the fraction of papers that are
  cited at least once more than X years after they are published, with
  a linear regression overlay.}
\label{fig:pldi-survival}
\end{figure}

We designed a series of visualizations to analyze citation patterns as
a function of venue in more
depth. Figures~\ref{fig:pldi-in-out} and ~\ref{fig:pldi-survival}
shows two example `flowers'  for PLDI.
Flowers for a wide range of
conferences are available on the ANU's Computational Media Lab's webpage~\cite{web:anu-cml-2019}.
More motivation and descriptions of these visualizations are
in a separate paper~\cite{shin2019influence}.
Because ACM
citation data
is limited to ACM DL venues, these visualizations use data from the Microsoft Academic Graph
(MAG)~\cite{sinha2015overview}. MAG curates a much larger corpus of
computer science venues than ACM and it provides an open-access
querying and analysis of all their citation
data.  

Figure~\ref{fig:pldi-in-out} visualizes the amount of
incoming and outgoing citations between PLDI and 25 other venues. The
plot contains a union of 25 venues that cite PLDI papers the most, or
{\em outgoing} scientific ideas from PLDI (in red), and top 25 cited
by PLDI papers, or {\em incoming} scientific ideas (in blue). We place
PLDI at bottom center, and place the other venues, including both
journals and conferences on a half-circle. Edge width represent the
volume of citations in either direction. Nodes are sized by the total
volume of citations in either direction, and ordered (from right to
left, blue to red) by ascending ratio of outgoing ideas to incoming
ideas. In other words, the six blue-ish nodes such as Communications of
the ACM are cited more by PLDI than they cite PLDI; and the 25 red-ish nodes
cite PLDI more than PLDI cites them.

The incoming/outgoing flow to PLDI itself, by definition, are equal,
and of high volume (thick edges). We can see that PLDI, focusing on
programming language design and implementation, is influenced by
broad-scope computer science journals such as Communications of the
ACM and conferences such as Operating System Design and Implementation (OSDI). PLDI in turn influences
conference such as ASPLOS (International Conference on
Architectural Support for Programming Languages and Operating Systems)
and the more specialized ECOOP conference (the European Conference on
Object-Oriented Programming).

Figure~\ref{fig:pldi-survival} describes {\em citation survival} over
time --- the fraction of papers (that are at least X years old) that
are cited at least once more than X years after being published. We
can see that about 90\% (at x=0) of PLDI papers are cited at least
once, and 60\% of papers are still being cited after 20
years. Quantifying the long-term impact of papers and venues has been
of interest to the computer science and science communities alike~\cite{wang2013quantifying}. Citation survival graph is an  intuitive visualization tool for this purpose.

Such in-depth analysis of citation statistics for a venue provides a view of scholarly impact that complements the ones based on authors and organizations. These graphs visualize the flow of ideas among communities, quantify the pace of innovation or the last impact of a scientific community, and trace the interaction among different sub-disciplines.

Just as any data sourced from the world-wide web, the data quality on
citations is not perfect. While the number of PLDI papers found by the
MAG mostly agree with the statistics at the publisher, for some venues
there are missing or spurious paper entries. Such data issues can
arise from data recording ambiguity at publishers, conference (or
journal) venue resolution, publishing of the same or similar papers
with varying detail in different venues, and a variety of other
sources. While we recommend that the academic community embrace quantitative tools for visualizing and analyzing scientific impact, we caution the reader to take online statistics with a grain of salt, and validate them with well-known community know-how when such knowledge exists.

\section{Top Department Behavior}

This section further explores the relationship between quality and
productivity by examining practices at four top departments.  Clauset
et al.\/ observe that faculty hiring is an assessment of research and
training quality that is widely distributed and fundamentally shapes
academic disciplines~\cite{CAL:2015}. Across computer science,
business, and history, they find doctoral prestige predicts faculty
hiring, reflecting
social inequality in academia. For instance, department prestige
predicts PhD production, increasing social advantage, and women place
worse than men even with a degree from the same program.  Clauset et
al.\/ identify Stanford (1), Berkeley (2), MIT (3), and CMU (7) as
dominant producers of computer science faculty between May 2011 and
March 2012. Using Taulbee data for 2003 and 2014 with current PhD
granting institutions of the faculty at 104 departments, we select
these four departments. We correlate these quality judgements to
productivity.   We find researchers at four top departments are neither
more nor less productive than all authors in the ACM, except for the
very most productive established authors at these four institutions.

We find researchers at four top departments are much more
collaborative than other researchers with 3.92 average authors per
paper in 2012, compared to 3.37 for all of ACM. Using the Student
T-test, we compare the average number of authors per paper from all of
ACM with papers from top departments in 2012 and obtain a $p$-value
$<< 0.0001$.  Figure~\ref{fig:acm-avg-authors} shows the trends over
time --- researchers at four top have become increasingly more
collaborative over time.  Deeper bibliometrics analysis shows that
collaboration networks are deeply coupled with citation networks and
reinforce scientific influence~\cite{BMG:2004}.

We illustrate these networks with the Taulbee data, which holds rank
constant over time for purposes of longitudinal
analysis~\cite{Taulbee:2015}.  Based on faculty hiring, North American
computing departments support the Taulbee ranking as shown in
Table~\ref{fig:faculty-origins}. Each column in the table presents the
percentage of faculty that earned their PhD at one of the top 4, 10,
15, and 25 departments and is now on the faculty of the row ranked
department.\footnote{This Taulbee Survey data was collated by Charles
  Isbell~\cite{elitism:2015} to demonstrate that if the top-ranked
  departments produced diverse PhDs, it would have a disproportionate
  influence on faculty and industry diversity in computing.} Hiring
reflects this ranking since a disproportionate amount, 43\% of
tenure-track faculty at the 25 top ranked U.S. departments, earned
their PhDs from one of these four departments. Furthermore, 78\% of
faculty at the top 25 institutions earned their PhDs from a top 25
institution. Combining this data with Table~\ref{fig:faculty-phd},
which reports raw number of Faculty and PhD production, we note that
these four departments employ 7.9\% of all faculty at the reporting
104 departments and produce 11.5\% of PhDs.  Although this higher PhD
production supports overall hiring trends, hiring of top four PhDs
into all Northern American faculty positions is much higher than
production. Clauset et al. disprove the hypothesis that quantity of
PhD production predicts placement~\cite{CAL:2015}. The academic
community hiring practices thus judge PhDs from these top four
departments as high quality.

\begin{figure}

\includegraphics[width=.9\columnwidth]{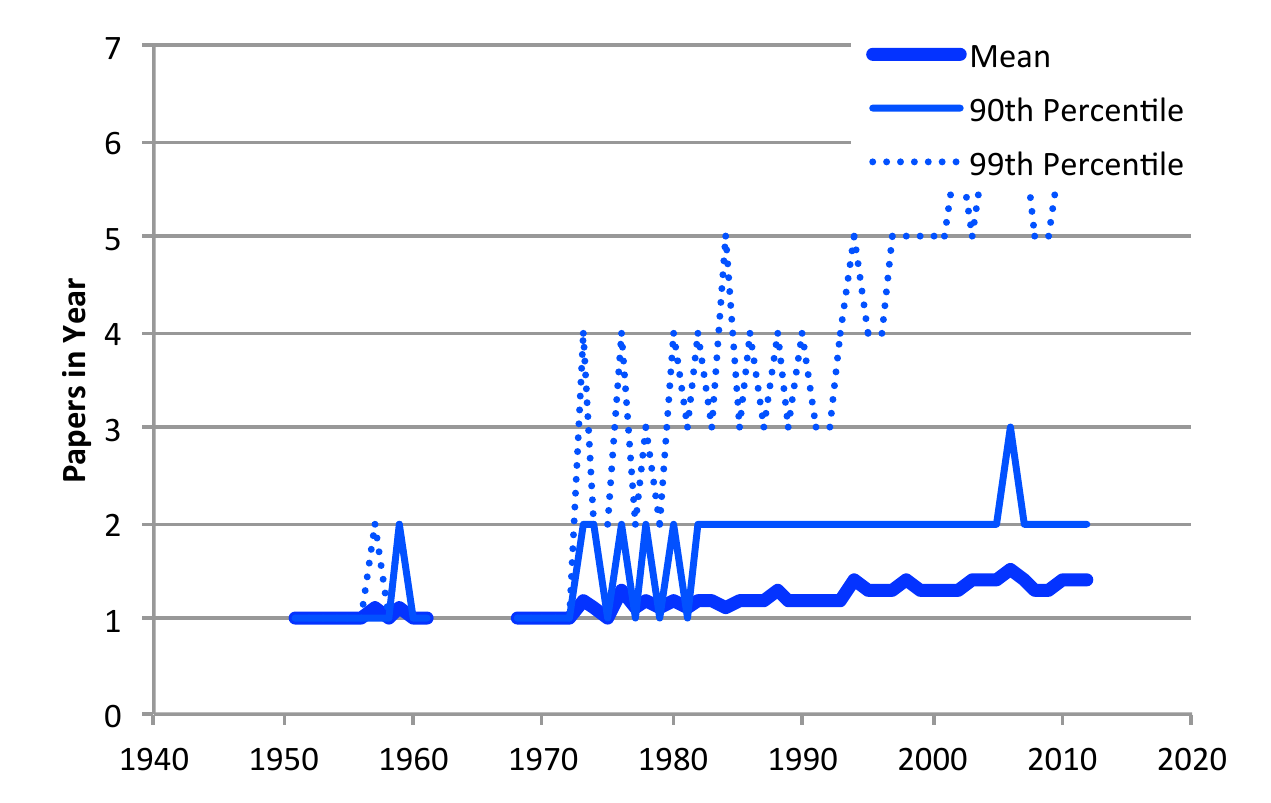}
\caption{Raw publications per year per top four department
  author. Same trend as all ACM authors.}
\label{fig:acm-raw-papers-per-top-author}

\includegraphics[width=.9\columnwidth]{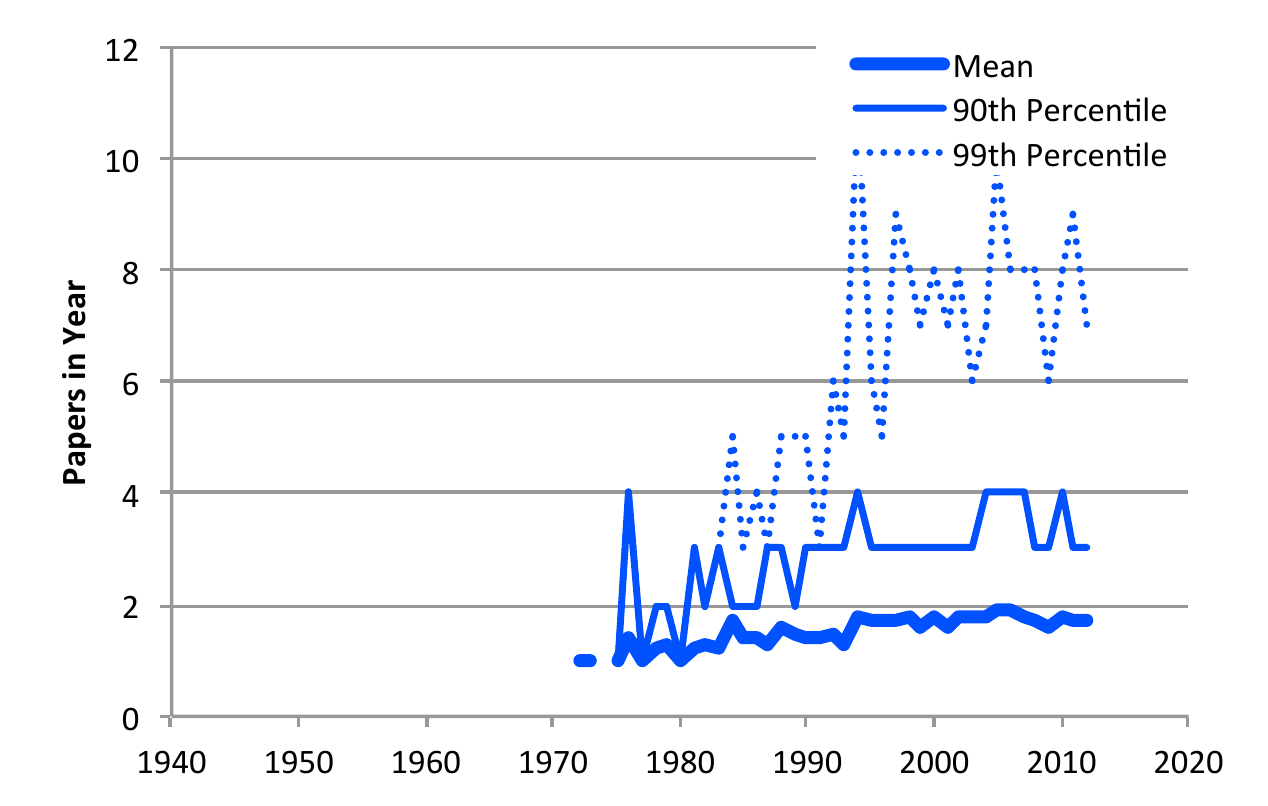}
\caption{Raw publications per established   ($N_e=5$)  author from one
  of the
  four top departments.  Established authors from four top departments produce more papers than other
  established authors.}
\label{fig:acm-raw-papers-per-established-top-author}

\includegraphics[width=.9\columnwidth]{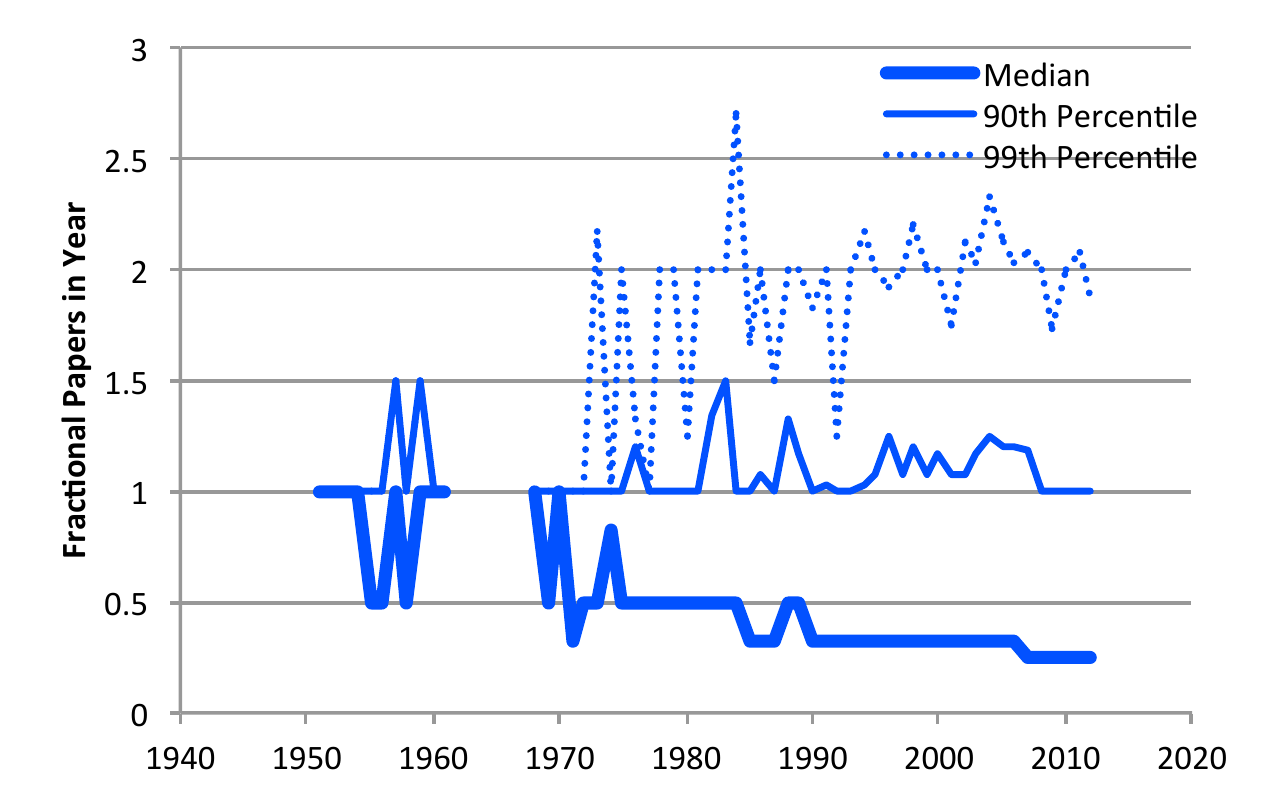}
\caption{Weighted publications per year per four top department
  author are less than all  ACM authors. }
\label{fig:acm-fraction-papers-per-top-author}

\includegraphics[width=.9\columnwidth]{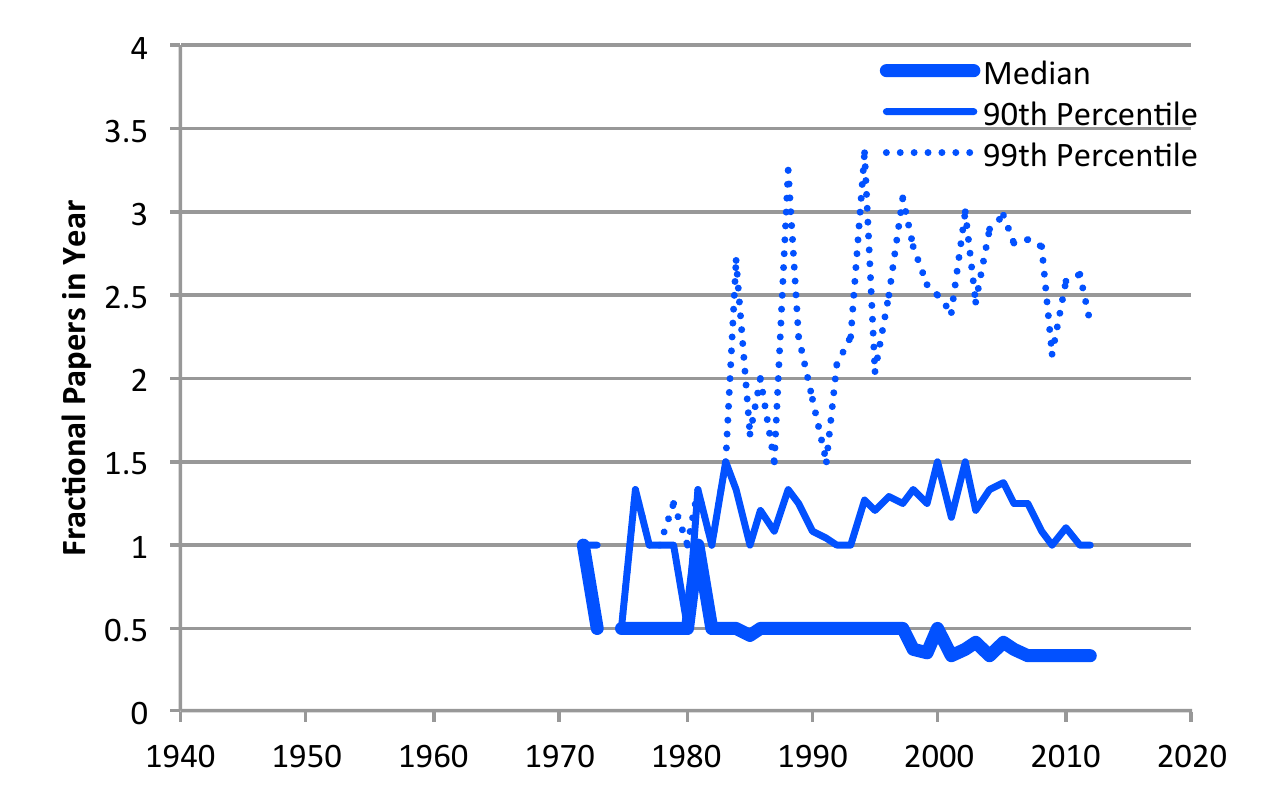}
\caption{Weighted publications per established  ($N_e=5$) top four author are
  similar to all established ACM authors}
\label{fig:acm-fraction-papers-per-established-top-author}

\end{figure}

Now we correlate U.S. faculty and PhD quality with publication and
author growth. 
In terms of U.S. PhD production, even though total PhD production grew
faster at 7\% per annum, four top departments continued to
disproportionally produce PhDs --- 11.5\% of PhDs graduate from these
departments.  Authors in these four departments are growing less than
the field as a whole: 9.7\% per annum versus 10.3\%. Reconsidering
Figure~\ref{fig:acm-institutions}, they produce disproportionate
amounts of research compared to other institutions, but the amount is
declining. Figures~\ref{fig:acm-raw-papers-per-top-author}
to~\ref{fig:acm-fraction-papers-per-established-top-author} show raw
papers from these four departments, raw papers from established
($N_e=5$) top four, weighted papers, and weighted papers from
established top four authors.  These 252 established researchers
produced 1587 publications in 2012, the average top department
established researcher has 6.3 papers, putting them in the 99th
percentile of all and established authors.  Note that the 90th
percentile of established top four authors produces four or fewer
papers a year, whereas the 90th percentile of all established authors
produce three or fewer. At the 99th percentile, authors from top
departements produce more than others, but otherwise have similar
productivity. As shown in Figure~\ref{fig:acm-avg-authors}, their
productivity stems from collaboration --- they co-author papers
with more authors.

Top departments set quality and production standards in several ways.
First, a plurality of all faculty were trained at these four
departments.  Second, others emulate their behaviors because these
researchers are considered among the very best. While most produce at
least two papers, the most prolific 10\% produce four or more
publications per year. 




\begin{table}
\footnotesize
\begin{center}
\begin{tabular}{@{}r|rrrr@{}}
\multicolumn{1}{c}{\textsf{\textbf{}}}
& \multicolumn{4}{c}{\textsf{\textbf{Rank of}}}\\
\multicolumn{1}{c}{\textsf{\textbf{Rank of}} }
& \multicolumn{4}{c}{\textsf{\textbf{PhD Granting Institution}}} \\
\multicolumn{1}{@{}c}{\textsf{\textbf{Faculty Institution}}}
&\textsf{\textbf{ $\leq$ 4}} & \textsf{\textbf{$\leq$ 10}} & \textsf{\textbf{$\leq$ 15}}  & \textsf{\textbf{$\leq$ 25}} \\ \midrule
\textsf{\textbf{$\leq$ 4}} &  59 & 67 & 75 & 82 \\
\textsf{\textbf{$\leq$ 10}} & 49 & 63 & 71 & 79 \\
\textsf{\textbf{$\leq$ 15}} & 46 & 62 & 72 & 80 \\
\textsf{\textbf{$\leq$ 25}} & 43 & 59 & 67 & 78
\end{tabular}
\end{center}
\caption{\textsf{\textbf{Origins of North American Tenure Track Faculty}} (2014) as a
  function of Taulbee
  PhD institution rank~\cite{elitism:2015}. Top ranked departments hire
  disproportionately from other top ranked departments.}~\\
\label{fig:faculty-origins}
 \end{table}

 \section{Related work}

 We next put these results in the context of two similar subdiscipline
 studies of publication practices for security and software
 engineering~\cite{Balzarotti:15,SE:Healthy:2014}. Balzarotti
 analyzes four top system security conferences between 2005 to 2015:
 Oakland, CCS, USENIX Security, and NDSS~\cite{Balzarotti:15}. The
 analysis includes submissions, publications, authors, affiliations,
 and nationality. The overall findings are consistent with our analysis
 on publication growth, author growth, established authors, and
 participating institutions (growth in international authors and
 collaborations).  The per annum author growth rate is 6.5\% for these
 venues, below what we compute for the entire field, but higher than
 the sample of top conferences in Table~\ref{tab:conferences}.  The
 acceptance rates have remained relatively steady around 15\%, thus
 these conferences remain similarly competitive. Another finding of
 note is that although 600 institutions authored papers in the security
 publications study, the top 10\% of institutions are responsible for
 77\% of the papers from 2005 to 2015.

 For software engineering, Vasilescu et al.~\cite{SE:Healthy:2014}
 study authors, program committees, and publications from 1994 to 2012
 for nine ACM and IEEE conferences. Because they hold the venues
 constant, their study is most similar to our conference data, but with
 a broader range of venue quality.  By following individual authors and
 program committee members over time, they compare conferences and
 analyze how the community changes over time.  ICSE (the most
 prestigious venue with the most citations, submissions, and lowest
 acceptance rate in their set) is increasingly \emph{less open} ---
 i.e., submissions are less likely to be accepted at ICSE when none of
 the authors has published in ICSE in one of the previous four
 years. One interesting note is that ICSE was not double blind in these
 years, which
 may be a root cause of this effect.
 On the other hand, the less prestigious ASE, FASE, and GPCE are
 \emph{more open} to new authors --- i.e., submissions are neither more
 or less likely to be accepted if authors published recently in these
 venues. \checked{These ICSE findings complement our results on the
   increase in collaboration and the higher increase at top venues,
   such as ICSE.}  The average paper at ICSE has both \emph{more ICSE
   experience} and \emph{more authors} than other SIGSE venues.

\section{Conclusions}

Since computer science is a relatively young discipline, many
researchers still remember developing or at least reading and
understanding, foundational results every year in a broad range of
computer science areas.
Today researchers struggle to read the significant publications in
their area, much less in all of computing.  However, computing is
continuing to expand and diversify --- new SIGs and their venues
emerged to explore promising research directions and many encouraged
more holistic research.  We believe that the community should embrace this growth as an indicator
that our field is vibrant.

The field is likely to keep growing.  The U.S.\/ Department of Labor
Statistics predicts increasing demand for computer
scientists~\cite{Jobs}. Undergraduate enrollments are booming.  Many
new education initiatives in primary school (K-12)  education are adding
computer science and computational education to their curriculums,
better preparing and interesting students in computing. These trends
presage 
continued growth in participants, venues, and publications in
computing research.

How to best handle high quality publication growth, reviewer overload,
and deliver expert comprehensive reviews to submissions remain very
important open questions.  \opinion{We believe that there is no reason
  to abandon proven best scientific practices, such as thorough expert
  reviewing, just because the field is growing. Our community needs to
  embrace growth and plan for it, creating more scalable reviewing and
  publication models.}  Many communities are now using tiered
reviewing with a large group of external reviewers and a smaller set
of program committee members to deliver expert reviews and spread
reviewer load. The program committee attends in-person PC meetings to
make final decisions, where we believe important community values are
shared and
developed~\cite{web:kathryn:2015,steve:pldi-chair:2015}. Another
benefit of the computer science conference system is that the program
committee changes every year and makes decisions with many witnesses,
distributing power and reducing factionalism. Interesting questions
for future analysis include (i) tracking submissions, rejections, and
final venue quality; (ii) correlations between program committee
membership, acceptance rates, and citations; and (iii) differences
between ACM journal and conference practices.  Our work invites more
in depth treatment of these topics, reviewing quality, and conference organization.

Since publication quality is correlated with collaboration,
\opinion{we expect increasingly more researchers and especially
  pre-tenure faculty will need to produce collaborative research and
  publications to be successful~\cite{teams:2015}.} Collaboration
requires sharing and rotating of research leadership and
responsibilities. If PhD students, post doctoral students, or
pre-tenure faculty were to abandon supporting research roles with
peers or senior faculty, they would miss opportunities to learn from
others. Furthermore, departments would miss the rich technical
cross-pollination that is sparked by the arrival of new researchers.
\opinion{Hiring and promotion committees should therefore encourage
  collaboration, rather than giving credit to only one author in a
  collaboration or worse ignoring highly collaborative impactful
  research altogether.}  Another point worth mentioning is that if
only the researchers at lower ranked departments were to decrease
their output (based on the false claim that quality is
correlated with less output) and top departments were to maintain their current
effective practices, researchers at lower ranked departments would be
further disadvantaged when compared to top department standards.

\section*{Acknowledgements}
We thank David Hawking, Kuansan Wang, David Walker, Susan Davidson,
and Erik Altman for discussions and suggestions on earlier drafts.  We
thank ACM for all the data, and
Bernard Rous at ACM for his help. We thank Betsy Bizot at the
Computing Research Association (CRA) for the Taulbee data on faculty
and departments. Our summary
analysis is available as supplementary material on arXiv~\cite{BMX:ACM:analysis:2019}.

\bibliographystyle{abbrvnat}
\bibliography{publication-practices.bib}


\end{document}

%% file: sig-table.tex
\centering
\footnotesize
\begin{tabular}{@{}l@{}r@{}rrr@{}r@{}}
& \multicolumn{2}{c}{\textsf{\textbf{2012~~~~~}}} & \multicolumn{3}{c}{\textsf{\textbf{\% $\Delta$ per annum 1990 - 2012}}} \\ 
\textsf{\textbf{SIG name}} &	\textsf{\textbf{Papers}} &
\textsf{\textbf{~~Avg. Authors}} &\textsf{\textbf{Papers}} &	\textsf{\textbf{Authors}} & \\ \midrule
SIGCHI & 	2562  &	3.49 & 15 & 16\\
SIGGRAPH &	1149 &  3.07 &	8 & 8\\
SIGWEB &	1017 & 3.35 &	17 & 17\\
SIGDA &	1002 & 3.73 &	7 & 8\\
SIGIR &	1000 & 3.39 &	18 & 18\\
SIGSOFT &	917 & 3.22 &	9 & 12\\
SIGARCH &	827 & 4.00 &	6 & 8\\
SIGMOD &	743 & 3.56 &	11 & 13\\
SIGPLAN &	739 & 3.13 &	8 & 9\\
SIGMOBILE &	736 & 3.93 &	26 & 27 & \scriptsize{~~1996-2012}\\
SIGSIM &	692 & 3.04 &	6 & 8\\
SIGMM &	569 & 3.54 &	10 & 11 & \scriptsize{1994-2012}\\
SIGAI &	526 & 3.27 &	3 & 5\\
SIGAPP &	476 & 3.35 &	5 & 8   &     \scriptsize{1992-2012}\\
SIGCOMM &	470 & 4.16 &	13 & 16\\
SIGKDD &	431 & 3.51 &	15 & 17 & \scriptsize{1999-2012}\\
SIGBED &	425 & 3.96 &	28 & 29 & \scriptsize{2003-2012}\\
SIGCSE &	395 & 2.82 &	8 & 8\\
SIGEVO &	390 & 2.75 &	2 & 2 & \scriptsize{2005-2012}\\
SIGSAC &	377 & 3.38 &	19 & 16 & \scriptsize{1992-2012}\\
SIGACT &	363 & 2.66 &	0 & 1\\
SIGOPS &	356 & 3.97 &	7 & 10\\
SIGMETRICS &	309 & 3.85 &	10 & 12\\
SIGMICRO &	248 & 3.59 &	9 & 11\\
SIGSPATIAL &	203 & 3.47 &	19 & 19 & \scriptsize{2005-2012}\\
SIGBio &	122 & 3.88 &	-4 & -1 & \scriptsize{1994-2012}\\
SIGHPC &	105 & 5.56 &	NA & NA & \scriptsize{2012-2012} \\
SIGACCESS &	78 & 3.31 &	15 & 13 & \scriptsize{2004-2012}\\
SIGSAM &	78 & 2.08 &	-1 & 0 & \scriptsize{1990-2011}\\
SIGecom &	73 & 3.04 &	10 & 9 & \scriptsize{1999-2012}\\
SIGITE &	62 & 2.45 &	2 & 5  & \scriptsize{2003-2012}\\
SIGDOC &	57 & 2.42 &	5 & 7\\
SIGUCCS &	49 & 1.94 &	-2 & 0\\
SIGMIS &	36 & 2.53 &	-1 & 0\\
SIGAda &	22 & 1.91 &	-8 & -8\\
SIGCAS &	19 & 1.68 &	-4 & -5 & \scriptsize{1990-2008}\\
\midrule
\textsf{\textbf{4 Top Dept.}} & 1,587 & 3.92 & 8.4  & 7.5 & \\ 
\textsf{\textbf{All of ACM}} & 14,521 & 3.37 &  9.3 & 10.3 & \\ 
\end{tabular}

%% file: conference-table.tex
\footnotesize 
{\centering
\begin{tabular}{@{}rrrrr|rrcrr@{\,\,\,}r@{}}
&  \multicolumn{4}{c}{\textsf{\textbf{1990 -
      2012}}} \\ 
&  \multicolumn{4}{c}{\textsf{\textbf{\% per annum change \%}}} 
&  \multicolumn{2}{c}{\textsf{\textbf{2012}}}
& & \multicolumn{2}{c}{\textsf{\textbf{\% accept \% }}}\\
& \multicolumn{2}{c}{\textsf{\textbf{papers}}} & 
\multicolumn{2}{c|}{\textsf{\textbf{authors}}} &
\multicolumn{2}{c}{\textsf{\textbf{avg $|$authors$|$}}} 
& & \textsf{\textbf{double}} 
& \textsf{\textbf{2012}} & \textsf{\textbf{1990 - 2012}} \\
\textsf{\textbf{Conference}} 
& \textsf{\textbf{conf.}} & \textsf{\textbf{SIG}} 
& \textsf{\textbf{conf.}} & \textsf{\textbf{SIG}} 
& \textsf{\textbf{conf.}} & \textsf{\textbf{SIG}} 
& \textsf{\textbf{$p$}} 
& \textsf{\textbf{blind?}} 
& \textsf{\textbf{rate}} & \textsf{\textbf{change}}\\ \midrule
CHI	& 8.2	& 11.0	& 9.9	& 12.4	& 3.7	& 3.5	         & 0.007 & yes & 23 & 0.000\\
SIGGRAPH\mbox{*}\hspace{-.52em}	& 3.6	& 8.5	& 6.0	& 9.2	& 3.6	& 3.1 & 0.002 & no & 19	& 0.000\\
WSDM	& 27.6	& 18.5	& 25.2	& 19.6	& 3.6	& 3.4         & 0.000 & no &  21	& 0.015\\
DAC	& 2.6	& 7.3	& 4.3	& 8.5	& 3.9	& 3.7	         & 0.069 & yes & 23 & -0.008\\
SIGIR	& 6.2	& 13.3	& 8.8	& 14.1	& 3.6	& 3.4         & 0.342 & yes & 20	& -0.004\\
ICSE	& 5.8	& 12.0	& 7.7	& 12.8	& 3.6	& 3.2                & 0.058 & no & 21	& -0.004\\
ISCA	& 1.3	& 5.3	& 2.9	& 7.4	& 4.1	& 4.0        & 0.824 & yes & 18	& 0.001\\
SIGMOD	& 4.5	& 9.7	& 6.6	& 11.6	& 4.2	& 3.6         & 0.049 & yes & 17	& 0.003\\
PLDI	& 2.2	& 7.0	& 5.2	& 8.1	& 3.9	& 3.1 	          & 0.000 & yes & 19 & -0.001\\
MOBICOM	& 2.7	& 23.7	& 5.7	& 25.0	& 4.7	& 3.9 & 0.008 & yes &	15 & -0.005\\ \midrule
\textsf{\textbf{Median}}	& 4.1	& 10.4	& 6.3	& 12.9 & 3.8 & 3.5 & & & 	& -0.001\\
\textsf{\textbf{All of ACM}}  & \multicolumn{2}{c}{9.3} &
\multicolumn{2}{c|}{10.6} & \multicolumn{2}{c}{3.4}\\
\end{tabular}}